\documentclass[10pt,journal]{IEEEtran} 

\usepackage{cite}
\usepackage{hyperref}
\usepackage{url}
\usepackage{tikz}
\usepackage{graphicx}
\usepackage[caption=false]{subfig} 
\usepackage{float}
\usepackage{stfloats}  
\usepackage{threeparttable}
\usepackage{booktabs}
\usepackage{multirow}
\usepackage{makecell}
\usepackage{caption}  
\usepackage{diagbox} 
\usepackage{array}
\usepackage[table]{xcolor} 
\usepackage{xcolor}
\usepackage{caption}

\usepackage[cmex10]{amsmath} 
\usepackage{amssymb}  
\usepackage{bm}       
\usepackage{mathrsfs} 
\usepackage{amsthm}   

\usepackage[linesnumbered,ruled,vlined]{algorithm2e}
\usepackage{xcolor}      

\usepackage{array}
\usepackage{mdwmath}     
\usepackage{mdwtab}      
\usepackage{eqparbox}    
\usepackage{etoolbox}    
\usepackage{supertabular} 

\usepackage{flushend}

\newtheorem{remark}{Remark}

\newtheorem{prop}{Proposition}

\usepackage[acronym]{glossaries}
\newacronym{6g}{6G}{sixth-generation}
\newacronym{5g}{5G}{fifth-generation}
\newacronym{iot}{IoT}{Internet-of-Things}
\newacronym{aiot}{AIoT}{artificial intelligence-of-things}
\newacronym{ai}{AI}{artificial intelligence}
\newacronym{mllm}{MLLM}{multimodal large language model}

\newcommand{\e}{\begin{equation}}
\newcommand{\ee}{\end{equation}}
\newcommand{\eqn}{\begin{eqnarray}}
\newcommand{\eeqn}{\end{eqnarray}}
\usepackage{balance}

\begin{document}

\title{ToDMA: Large Model-Driven Massive Token Communications for Semantic Multiple Access}

\author{Li Qiao,~\IEEEmembership{Member,~IEEE}, Mahdi Boloursaz Mashhadi,~\IEEEmembership{Senior Member,~IEEE}, \\Zhen Gao,~\IEEEmembership{Senoir Member,~IEEE}, Robert Schober,~\IEEEmembership{Fellow,~IEEE}, and Deniz Gündüz,~\IEEEmembership{Fellow,~IEEE}
 \vspace{-8mm}

\thanks{Part of the work was accepted by IEEE INFOCOM 2025 Workshop \cite{Qiao2025Token}. L. Qiao is with The University of Hong Kong, Hong Kong (e-mail: qiaoli@hku.hk). Z. Gao is with Beijing Institute of Technology, Beijing 100081, China (e-mail: gaozhen16@bit.edu.cn). M. Boloursaz Mashhadi is with 5GIC \& 6GIC, Institute for Communication Systems (ICS), University of Surrey, GU2 7XH Guildford, U.K. (email: {m.boloursazmashhadi}@surrey.ac.uk). Robert Schober is with the Institute for Digital Communications, Friedrich-Alexander-University Erlangen-Nurnberg, 91054 Erlangen, Germany (e-mail: robert.schober@fau.de). D. Gündüz is with the Department of Electrical and Electronic Engineering, Imperial College London, London SW7 2AZ, U.K. (email: d.gunduz@imperial.ac.uk).}
\thanks{D. Gunduz received funding from UKRI for the project AI-R (ERC Consolidator Grant, EP/X030806/1) and the SNS JU project 6G-GOALS under the EU’s Horizon program (Grant Agreement No. 101139232).}
}
\maketitle

\vspace{-5mm}
\begin{abstract}
Token communications (TokenCom) is an emerging generative semantic communication paradigm, where tokens serve as compact representation units across modalities. Their contextual dependencies can be exploited by pretrained large models for semantic recovery. In this paper, we propose token-domain multiple access (ToDMA), a large-model-driven semantic multiple access scheme for massive token communications. ToDMA integrates unsourced random access with context-aware token processing. It enables massive uncoordinated devices to transmit tokenized source representations over common uplink resources. Specifically, each token index is associated with a shared modulation codeword, exposing token-level structure to the receiver for context-aware recovery. At the receiver, compressed sensing is first employed to jointly detect active tokens and estimate their corresponding channel state information (CSI) from the superposed signals. The source token sequences are then reconstructed by exploiting the consistency of token-associated CSI across multiple token positions. In the presence of token collisions, some active tokens may remain unassigned, leading to missing entries in the reconstructed token sequences. To recover these tokens, candidate-restricted masked-token prediction is performed using pretrained contextual models, thereby leveraging token-level context to mitigate collision effects. Simulation results on both image and text transmission tasks demonstrate
that ToDMA reduces access latency while maintaining favorable token recovery and semantic reconstruction quality, showing its scalability for semantic multiple access.

\end{abstract}
\begin{IEEEkeywords}
Token communications, generative semantic communications, massive communication, unsourced random access, context-aware token recovery.
\end{IEEEkeywords}

\IEEEpeerreviewmaketitle
\section{Introduction}\label{C:Intro}
The rise of \glspl*{mllm} marks a significant breakthrough in artificial intelligence (AI), extending large language models (LLMs) to process and reason across multiple modalities, such as text, images, video, and audio \cite{Duzhen2024}. Recent \gls*{mllm}s, such as BLIP-2 and LLaVA, exhibit strong cross-modal reasoning, enabling tasks like visual question answering and multimodal content generation \cite{li2023blip,liu2024visual}. These advances are expected to profoundly impact next-generation wireless networks by enabling generative semantic communications, where the goal shifts from delivering raw bits to conveying meaningful content with improved efficiency and robustness \cite{lu2024generative,jiang2024large,   Liang2024generative}.

\subsection{Tokens: The Currency of Generative AI}
A \textit{token} in generative AI refers to the unit of information a model processes. For text, a token may correspond to a word fragment or a group of characters; for images, it tipically represents image patches or features; for audio, it can be a sound segment. Across modalities, tokens serve as discrete units that enable models to capture underlying patterns, context, and relationships.
At the core of \gls*{mllm}s is the \textit{transformer} architecture, which is designed to efficiently process {\it tokens}—whether they represent words, patches of images, or other forms of data \cite{vaswani2017attention}. Transformers utilize self-attention mechanisms to capture relationships between tokens, regardless of their modality, allowing \gls*{mllm}s to comprehend both the context of language and the semantics of visual or auditory inputs \cite{Duzhen2024}. 

The process of transforming signals into tokens is called \textit{tokenization}. This involves utilizing a \textit{tokenizer}, typically a learned codebook, to map segments of the input signal to discrete tokens. Tokenization allows complex data to be broken into manageable and interpretable units. For instance, the well-known BERT model employs a WordPiece-based tokenizer that has around 30,000 tokens to process natural language \cite{devlin2018bert}. The sentence ``I love transformers!" is tokenized as [[CLS], I, love, transform, \#\#ers, !, [SEP]], with the corresponding token IDs being [101, 146, 1567, 11303, 1468, 106, 102]. Each token ID corresponds to a vector embedding, which is then fed into transformer models for further processing. Another example is the vector quantized variational autoencoder (VQ-VAE) tokenizer for images \cite{van2017neural}. The VQ-VAE tokenizer converts continuous data (e.g., images, audio) to compact and discrete tokens by mapping it to a learned codebook of predefined vectors. Interested readers are referred to \cite{qian2022makes} for a comprehensive overview of vision tokenizers.

In the pre-training phase of generative transformer-based models, self-supervised training on a vast dataset containing billions of tokens is employed. A major category of such models like BERT \cite{devlin2018bert} are based on masked language modeling, in which random tokens are masked during pre-training, and the model learns to generate the masked tokens from the token codebook based on the surrounding semantic context, comprising available preceding and succeeding tokens. The prediction of masked tokens is framed as a classification problem, with pre-trained transformer models estimating the probability distribution of the masked token over the token codebook, based on the bidirectional context information. Masked image modeling has also been adopted in vision pre-training, e.g., MaskGIT \cite{chang2022maskgit}, yielding impressive results for various vision tasks. The {\it next-token prediction} approach is another effective pre-training task, as evident from the GPT models \cite{brown2020language} for text generation and the use of auto-regressive models for image generation \cite{esser2021taming}.
\subsection{What is Token Communication?}

\textit{Token communications (TokenCom)} was recently introduced in \cite{Magzine2025Token} as a novel framework that leverages tokens as units of semantic content transmitted within future wireless networks. Tokens provide compact representations of multimodal messages and exhibit contextual dependencies that can be exploited at both the transmitter and the receiver \cite{lee2025low,Qiao2025Token}. The transformer-based token processing in TokenCom enables the encoding of the semantic content of multimodal signals at the transmitter and recovery of the corresponding semantics at the receiver leveraging context information. For example, a corrupted TokenCom packet may result in an incomplete message at the receiver, such as ``A beach with palm [MASK] and clear blue water." Here, the missing word is represented by a [MASK] token. The semantic error correction mechanism in TokenCom leverages a pre-trained LLM to infer the masked word from the surrounding context, e.g., predicting ``trees" to recover the sentence as ``A beach with palm trees and clear blue water." This avoids packet retransmissions that would otherwise be required without exploiting contextual information via the LLM \cite{Magzine2025Token}. This context-aware semantic recovery is particularly attractive for wireless networks with limited radio resources, as it can trade additional
receiver-side computation for reduced retransmission overhead.

\begin{figure*}[t]
     \centering
     \includegraphics[width = 1.8\columnwidth,keepaspectratio]{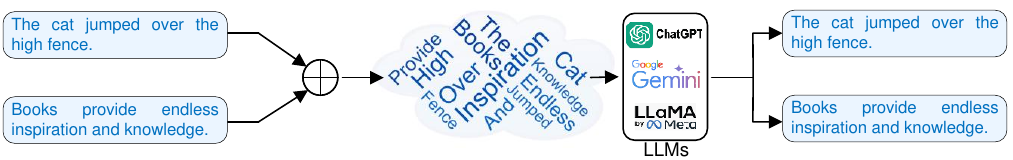}
     \captionsetup{font={footnotesize, color = {black}}, singlelinecheck = off, justification = raggedright,name={Fig.},labelsep=period}
     \caption{A conceptual illustration of {\it Semantic Orthogonality}: illustrating how contextual semantic differences between two text sources can be exploited by LLMs to reconstruct them from a mixed signal.}
     \label{fig1:LLMsMA}
\end{figure*}

\subsection{Token Domain Multiple Access (ToDMA)}

This paper focuses on uplink token transmission from a massive number of devices. We ask the following question: how should tokens, rather than bits, be modulated and transmitted over a multiple access channel (MAC)? A straightforward approach is to serialize token indices into bits and apply conventional channel coding and modulation. While feasible, this bit-domain approach does not directly expose the token codebook structure or contextual dependencies to the multiple access receiver. To address this issue, we propose a context-aware token-domain multiple access (ToDMA) scheme, where token indices are directly associated with shared modulation codewords. This allows massive uncoordinated devices to transmit token sequences over common uplink resources in a grant-free and non-orthogonal manner. More importantly, ToDMA establishes a token-level interface between physical-layer signal recovery and contextual semantic recovery. When multiple tokens cannot be uniquely separated at the physical layer, the remaining uncertainty can be further resolved by exploiting contextual dependencies.

We further introduce token-domain {\it semantic orthogonality} to characterize the contextual separability of physically ambiguous token candidates. Intuitively, two token sequences are more semantically orthogonal if their surrounding contexts allow a pretrained model to distinguish their collided tokens with high confidence. For instance, the sentences ``The cat jumped over the high fence'' and ``Books provide endless inspiration and knowledge'' provide substantially different contexts, making their ambiguous tokens easier to separate, as illustrated in Fig.~\ref{fig1:LLMsMA}. In contrast, if the ambiguous tokens correspond to context-insensitive words such as ``She'' and ``He'' in the sentences ``She likes to read novels'' and ``He dislikes to read novels'', the limited surrounding context may not provide sufficient information to determine the correct token. This observation highlights the need to combine physical-layer token detection with context-based token recovery, rather than relying on contextual prediction alone. A mathematical interpretation of token-domain semantic orthogonality is provided in Section~\ref{Sec:SO}, and its cross-modal validation is presented in Section~\ref{Sec:SO-sim}.

\section{Related Works}
\subsection{Generative AI Meets Semantic Communications}
Goal-oriented and semantic communications have been shown to possess significant potential for transforming next-generation wireless networks, making them more efficient, timely, and intelligent \cite{strinati2024goal,beyondBits}. These advancements primarily stem from AI-driven intent-aware joint optimization of source encoding, channel coding, modulation, etc. For example, deep joint source-channel coding (DJSCC) optimizes the source and channel coding simultaneously using autoencoders, thereby reducing the communication bandwidth while ensuring gradual quality degradation as channel conditions worsen for wireless multi-media transmission \cite{ Zhijin,bourtsoulatze2019deep}. With advancements in generative AI (GenAI), semantic communications can be further enhanced by utilizing generative models to improve the comprehension and generation of high-fidelity communication content, a concept recently proposed as {\it generative semantic communications} \cite{jiang2024large,Liang2024generative}.

GenAI has been shown to have significant potential to enhance semantic communications across various dimensions, including source coding and joint source-channel coding (JSCC) \cite{chen2024information}.
{\bf GenAI for Source Coding:} LLMs have been utilized for compression tasks, demonstrating superior performance as general-purpose compressors compared to domain-specific methods \cite{ deletang2023language}. Advances in vision generative models, such as generative adversarial networks (GANs) and diffusion models, enable ultra-low bit-rate transmission of semantic modalities (e.g., prompts, edge maps, embeddings), allowing the receiver to reconstruct high-quality signals \cite{qiao2024latency, cicchetti2024language, ExCompres, Yin25Video}. Moreover, intent-aware semantic decomposition aligns transmitted information with the receiver's goal \cite{liu2024diffusion}, and mobile edge servers can assist in handling the corresponding GenAI computational demands \cite{xu2024accelerating, ren2024generative}. {\bf GenAI for JSCC:} GenAI-powered DJSCC models have been proposed in \cite{ erdemir2023generative,yilmaz2024high, zhang2025semantics,chen2025sing}, achieving enhanced distortion and perceptual performance compared to conventional DJSCC methods. Along this line, a task-oriented adaptive token selection scheme and a token merging scheme were introduced in \cite{devoto2024adaptive} and \cite{erak2025adaptive}, respectively, to reduce bandwidth consumption. Furthermore, the work in \cite{Magzine2025Token} demonstrates that token context can be exploited to predict lost packets, effectively avoiding re-transmissions and reducing latency. The major performance gains of generative semantic communication schemes primarily stem from two key aspects:
\begin{itemize}
\item {\it Learned knowledge-base (KB) for compression:} The KB includes components such as the encoder/decoder weights in DJSCC, and the tokenizer codebook in \gls*{mllm}s and VQ-VAE, which represent information in a compact and efficient manner.
\item {\it JSCC:} End-to-end training-based frameworks enhance the performance particularly for short block packages.
\end{itemize}
Despite the remarkable success of GenAI-based semantic communications in point-to-point transmission, research on multi-user scenarios remains in its infancy. In particular, the potential of leveraging transformer-based token processing and MLLMs to mitigate multiuser interference and support efficient massive communication remains an underexplored frontier.

\subsection{Evolution of Multiple Access in 6G Networks}
Multiple access techniques are advancing rapidly towards the sixth-generation (6G), where approaches targeting efficient random access for a massive number of uncoordinated devices characterized by sporadic activity and short data packets are particularly relevant. Among the key technologies enabling massive communications, NOMA and grant-free random access (GFRA) are anticipated to play a crucial role in meeting future demands \cite{clerckx2024multiple}. 

Traditionally, NOMA improves bandwidth efficiency by exploiting power-domain or code-domain non-orthogonality, allowing multiple users to share the same resources while employing successive interference cancellation (SIC) at the receiver to mitigate multi-user interference. Most existing work on semantic multiple access follows this philosophy. By integrating DJSCC with NOMA, these schemes employ end-to-end training to enable neural networks to learn interference cancellation, implicitly leveraging the ``semantic orthogonality" or separability of source signals \cite{yilmaz2023distributed, liang2024orthogonal, yilmaz2025learning, Wang2025SemanticNOMA}. 
While effective in small-scale scenarios, scaling NOMA-based semantic schemes to massive access remains challenging. Specifically, the necessity of user grouping and grant-based scheduling for high-density scenarios introduces significant handshake latency, hindering the scalability required for future intelligent networks.

To address the scalability and latency bottlenecks, grant-free random access has been widely adopted, where active devices transmit unique preassigned signatures to allow the base station (BS) to perform joint activity detection, channel estimation, and data detection via compressed sensing (CS) techniques \cite{ZhenMag, ke2020compressive}. To further accommodate an even larger number of devices, unsourced random access (URA) replaces user-specific preambles with a shared codebook   \cite{polyanskiy2017perspective}. In URA, the BS focuses on decoding messages from the shared codebook without necessarily identifying the transmitters at the physical layer \cite{liva2024unsourced,YongpengAsyn, URA_survey}. If needed, device identity can be embedded in the packet payload and resolved at higher layers. There are also scenarios where individual device identities are not essential, such
as over-the-air computation or event-alarm sensing, where the receiver is mainly interested in an aggregate value or the occurrence of an event \cite{liva2024unsourced, qiao2024massive}.

In general uplink multiple access, however, the receiver still needs to recover distinct messages from different active devices. The non-orthogonal nature of URA therefore leads to potential codeword collisions when multiple devices select identical entries from the shared codebook, representing a persistent challenge in massive access scenarios. Existing studies have addressed this issue by exploiting angular-domain sparsity in massive MIMO systems~\cite{YongpengUMA}, or by implementing collision detection mechanisms where retransmissions are triggered for unrecoverable packets~\cite{YongpengJSAC}. While effective, these physical-layer or retransmission-based approaches may encounter latency and resource limitations as the network density continues to scale. Moreover, existing URA schemes are mainly designed for bit-level packet transmission, where the token structure and contextual dependencies of the payload are not considered. As a result, unresolved codeword collisions are treated as packet-level failures rather than token-domain uncertainty that can be refined using context. To the best of our knowledge, URA has not been designed for
tokenized massive communications with context-aware recovery, which motivates the proposed ToDMA framework.

\section{Contributions}
Motivated by the above gap, we propose ToDMA, a token-domain semantic multiple-access framework that integrates grant-free URA with context-aware token processing. The main contributions are summarized as follows:
\begin{itemize} 
\item {\bf Communication-efficient ToDMA framework:}
The proposed ToDMA framework directly operates on token sequences rather than first converting token indices into bit streams. Specifically, each token index is associated with a shared random access modulation codeword, allowing massive uncoordinated devices to transmit over common uplink resources in a grant-free and non-orthogonal manner. This
design reduces signaling overhead while
preserving the token-domain structure for context-aware processing.

\item {\bf Context-aware receiver architecture:}
The proposed receiver proceeds in three steps, namely, \emph{active-token detection}, \emph{token assignment}, and \emph{masked-token prediction}. Specifically, we propose a CS-based algorithm to jointly identify active tokens and estimate channel state information (CSI), followed by a clustering-based token assignment. To recover unassigned tokens caused by collisions, we leverage bidirectional transformers for context-aware semantic recovery. The receiver then predicts missing or ambiguous tokens from the candidate set, thereby mitigating the impact of non-orthogonal interference.

\item {\bf Semantic orthogonality and scalability analysis:}
We introduce an entropy-based interpretation of token-domain semantic orthogonality to quantify the contextual separability among ambiguous token candidates. We further analyze the receiver complexity and scalability of ToDMA under massive access scenario, showing that the shared token codebook
does not scale with the number of active devices and that multi-antenna observations improve active token detection performance.

\item {\bf Cross-modal and system-level validation:}
We validate ToDMA on both image and text transmission tasks. Simulation results show that ToDMA improves token recovery and
semantic reconstruction quality under limited uplink resources, compared with representative
semantic and digital transmission baselines. Latency, complexity, and runtime
analyses further quantify the resulting communication-computation
trade-off.
\end{itemize}

\textit{Notation}: Boldface lowercase and uppercase symbols denote column vectors and matrices, respectively. For a matrix ${\bf A}$, ${\bf A}^T$, ${\bf A}^*$, ${\bf A}^H$, ${\left\| {\bf{A}} \right\|_F}$, $[{\bf{A}}]_{m,n}$ denote the transpose, conjugate, Hermitian transpose, Frobenius norm, and the $m$-th row and $n$-th column element of ${\bf{A}}$, respectively. $[{\bf{A}}]_{:,n}$ ($[{\bf{A}}]_{n,:}$) denotes the $n$-th column (row) of ${\bf A}$. $\| \cdot \|_p$ denotes the ${l_p}$ norm of its argument. The function supp\{$\cdot$\} returns the indices of the non-zero elements in a vector, and the indices of the nonzero rows in a matrix. $[K]$ denotes the set $\{1,...,K\}$. $|\mathcal{P}|$ denotes the cardinality of set $\mathcal{P}$, and ${\bf 0}_{m\times n}$ (${\bf 1}_{m\times n}$) is a matrix of all zeros (ones) with dimension ${m\times n}$. Finally, $\mathcal{CN}(x;\mu,\tau)$ denotes the complex Gaussian distribution with mean $\mu$ and variance $\tau$. 

\begin{figure*}[t]
     \centering
     \includegraphics[width = 1.85\columnwidth,keepaspectratio]{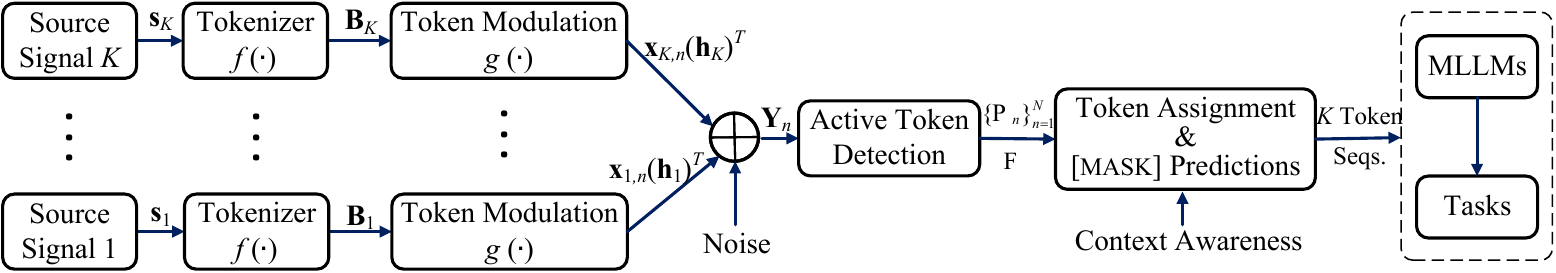}
     \captionsetup{font={footnotesize, color = {black}}, singlelinecheck = off, justification = raggedright,name={Fig.},labelsep=period}
     \caption{Proposed token-domain multiple access (ToDMA) framework.}
     \label{fig2}
     \vspace{-5mm}
\end{figure*}

\section{Proposed Token-Domain Multiple Access (ToDMA) Framework}

In this section, we first introduce the transmitter architecture for our proposed ToDMA framework. Then, we provide the corresponding signal model and problem formulation.
\subsection{Transmitter Design}
Fig. \ref{fig2} depicts the block diagram of the proposed ToDMA framework. We consider a multiple access scenario, where $K_{\rm T}$ single-antenna devices are served by a BS equipped with $M$ antennas. Despite a large number of devices, typically only a small fraction, $K$ out of $K_T$ (where $K \ll K_T$) have data to transmit at a given time \cite{liva2024unsourced}. We denote the source signal of the $k$-th active device as vector ${\bf s}_k$, $k\in[K]$, which can represent the data of any modality, like text, image, video, audio, etc., represented in a vector form. In the following, we will describe the tokenization and token modulation processes at each device.

\subsubsection{Tokenization}
A tokenizer partitions the original data vector ${\bf s}_k$ into little pieces, each belonging to a prescribed finite token alphabet. We note that tokenization is a paticular case of source coding. Eventually, the signal is mapped to a sequence of indices from this alphabet. 
We use $f(\cdot)$ to represent the tokenizer and $Q$ to represent the size of the {\it token codebook}. Therefore, the {\it tokenization} process can be represented by ${\bf B}_k = f({\bf s}_k) \in \{0, 1\}^{Q\times N}$, where the $n$-th column of ${\bf B}_k$, denoted by ${\bf b}_{k,n}$, is a one-hot vector representing the corresponding token, {with exactly one entry equal to one and all
remaining entries equal to zero.}. Without loss of generality, the length of the resulting token sequence $N$ is assumed to be the same for all devices. Later, we will use $f^{-1}(\cdot)$ to denote the de-tokenization process.

\subsubsection{Token Modulation}

Assuming that the same tokenizer is used by all devices, we propose to employ a common modulation codebook for multiple access, denoted as ${\bf U} \in \mathbb{C}^{L\times Q}$, where $L$ denotes the length of each codeword. For simplicity, here we generate the modulation codebook ${\bf U}$ by sampling each element from a complex standard Gaussian distribution. We use $g(\cdot)$ to represent the token modulation process; hence, the output is denoted by
${\bf X}_k  = g({\bf B}_k)= {\bf U}{\bf B}_k \in \mathbb{C}^{L\times N}$, where each token is modulated onto a codeword. We denote the $n$-th column of ${\bf X}_k$ as ${\bf x}_{k,n}= {\bf U}{\bf b}_{k,n} \in\mathbb{C}^{L}$. The proposed token modulation offers two key advantages. First, by employing a common modulation codebook, it supports uncoordinated random access from a massive number of devices, thereby reducing access latency and simplifying implementation. Second, it redefines the basic communication unit from conventional bits to semantic tokens, enabling the receiver to leverage transformer-based token processing and integration with \gls*{mllm}s.

\subsection{Multiple Access Signal Model}
In a typical massive communication scenario envisioned for future 6G networks, BSs serve a large number of uncoordinated devices \cite{liva2024unsourced}. Periodically, the BS broadcasts beacon signals. If a device has data to send, it transitions from idle to active state without the BS knowing which devices are active at any given time. Upon receiving the beacon signal, active devices immediately transmit over the MAC. We assume that $K$ active devices use the same $N$ communication resources, e.g., $N$ time slots, each comprising $L$ symbols, to transmit their $N$ individual tokens. In the $n$-th time slot, $\forall n\in[N]$, the transmitted signals overlap at the receiver, i.e., the received signal ${\bf Y}_{n}\in\mathbb{C}^{L\times M}$ can be expressed as
\begin{align}\label{MA-SignalMod}
    {\bf Y}_{n}&=\sum\nolimits_{k\in[K]} {\bf x}_{k,n} \left({\bf h}_k\right)^T + {\bf Z}_{n} \nonumber \\
    &={\bf U} \sum\nolimits_{k\in[K]}{\bf b}_{k,n} \left({\bf h}_k\right)^T + {\bf Z}_{n}= {\bf U}{\bf H}_n + {\bf Z}_{n},
\end{align}
where ${\bf h}_k\in \mathbb{C}^{M}$ denotes the channel vector between the $k$-th device and the BS, ${\bf H}_n = \sum\nolimits_{k\in[K]}{\bf b}_{k,n} \left({\bf h}_k\right)^T \in\mathbb{C}^{Q\times M}$ is referred to as the {\it equivalent channel matrix}, and the elements of the noise matrix ${\bf Z}_{n}$ follow independent and identically distributed (i.i.d.) complex Gaussian distributions with zero mean and variance $\sigma^2$. We assume a slow Rayleigh fading channel; that is, the elements of ${\bf h}_k$ are sampled independently from a complex standard Gaussian distribution, and are assumed to remain constant throughout the $N$ time slots.

{\it Row sparsity} of the {\it equivalent channel matrix}: Typically, the size of the token codebook is much larger than the number of active devices, i.e., $Q\gg K$. Recall that ${\bf b}_{k,n}$ is a one-hot vector. Thus, the row sparsity of ${\bf H}_n$ is obvious and can be expressed as
\begin{align}\label{RowSparse}
    |\text{supp}\{{\bf H}_n\}| \leq K \ll Q, ~~\forall n\in [N],
\end{align}
where the inequality holds as equality if and only if the tokens selected by the $K$ devices in a given time slot are all distinct, i.e., no token collision happens. 

\subsection{Problem Formulation}
To reconstruct the transmitted token sequences of the active devices, i.e., 
$\{{\bf B}_k, k\in[K]\}$, the receiver needs to recover not only the transmitted 
token indices in each time slot, but also their device-wise ordering across time. 
Accordingly, we formulate the receiver-side reconstruction task as three coupled 
subproblems: active token detection in each time slot, token assignment across 
time slots, and masked token prediction for unreliable or missing entries. 
These three subproblems are detailed as follows.

\subsubsection{Active Token Detection}
For each time slot $n\in[N]$, the 
active token detection can be formulated as the 
following row-sparse recovery problem
\begin{align}\label{subprob1}
    \min_{{\bf H}_n} \quad 
    \left\|{\bf Y}_n - {\bf U}{\bf H}_n\right\|_F^2,\quad \text{s.t.} \quad \text{(\ref{RowSparse})}.
\end{align}
We denote the true active token index set in the $n$-th time slot 
by $\mathcal P_n=\mathrm{supp}\{{\bf H}_n\}$, and its estimate by 
$\widehat{\mathcal P}_n$. By solving problem~(\ref{subprob1}), the receiver 
obtains an estimate of the equivalent channel matrix, denoted by 
$\widehat{\bf H}_n$, and the corresponding active token index set 
$\widehat{\mathcal P}_n$.

For each detected token index $\phi\in\widehat{\mathcal P}_n$, we define the 
estimated CSI as 
$\psi_n(\phi)=[\widehat{\bf H}_n]_{\phi,:}^{T}\in\mathbb{C}^{M}$. 
Accordingly, the set of all estimated CSI over $N$ time 
slots is given by 
$\widehat{\mathcal F}=\{\psi_n(\phi)\mid \phi\in\widehat{\mathcal P}_n, 
n\in[N]\}$. Since $L\ll Q$, especially when large AI models induce a large 
token vocabulary, problem~(\ref{subprob1}) is a challenging underdetermined 
sparse recovery problem.

\subsubsection{Token Assignment}
Define $\alpha_{k,n}^*=\mathrm{supp}\{{\bf b}_{k,n}\}$ and 
$\alpha_{k,n}=\mathrm{supp}\{\widehat{\bf b}_{k,n}\}$ as the true token index 
and the assigned token index of device $k$ in time slot $n$, respectively. 
After active token detection, the receiver obtains the detected token index sets 
$\{\widehat{\mathcal P}_n,n\in[N]\}$ separately for different time slots. 
However, these detected tokens are not yet associated with their corresponding 
devices. The goal of token assignment is to concatenate the detected tokens over 
time and form device-wise token sequences.

Ideally, the token assignment problem can be formulated as
\begin{align}\label{subprob2}
    \min_{\{\alpha_{k,n}\}_{k=1,n=1}^{K,N}} 
    &\sum_{k=1}^{K}\sum_{n=1}^{N} 
    \mathbf{1}\{\alpha_{k,n}\neq \alpha_{k,n}^*\}, \nonumber\\
    \text{s.t.}\quad 
    &\alpha_{k,n}\in\widehat{\mathcal P}_n,\quad \forall k,n,
\end{align}
where $\mathbf{1}\{\cdot\}$ is the indicator function. 
Problem~(\ref{subprob2}) is an oracle formulation, since the ground-truth token 
indices $\{\alpha_{k,n}^*\}$ are unavailable at the receiver.

To reveal the structure of this assignment problem, we next consider a 
channel-aided ideal formulation. Under the slow-fading channel assumption, the 
channel vector of each active device is approximately invariant over the $N$ 
time slots. Hence, the estimated CSI associated with the 
same device tend to concentrate around a common device-specific channel 
signature. If the channel vectors $\{{\bf h}_k\}_{k=1}^{K}$ were known, token 
assignment could be formulated as
\begin{align}\label{subprob2-2}
    \min_{\{\alpha_{k,n}\}_{k=1,n=1}^{K,N}} 
    &\sum_{k=1}^{K}\sum_{n=1}^{N} 
    \left\|\psi_n(\alpha_{k,n})-{\bf h}_k\right\|_2^2, \nonumber\\
    \text{s.t.}\quad 
    &\alpha_{k,n}\in\widehat{\mathcal P}_n,\quad \forall k,n.
\end{align}
However, the actual channel vectors $\{{\bf h}_k\}_{k=1}^{K}$ are unavailable 
at the receiver. Therefore, problem~(\ref{subprob2-2}) cannot be directly 
solved and serves only as a channel-aided ideal formulation. This observation 
motivates the clustering-based token assignment method developed in 
Section~\ref{Sec:TokAssignPre}, where the unknown device-specific channel 
signatures are inferred from the estimated CSI in 
$\widehat{\mathcal F}$.


\subsubsection{Masked Token Prediction}
After token assignment, the receiver obtains a preliminary estimate of the 
device-wise token sequences, denoted by $\{\widehat{\bf B}_k,k\in[K]\}$. 
Due to token collisions and possible errors introduced in the preceding active 
token detection and token assignment stages, some estimated tokens may be 
unreliable. To identify these unreliable entries, we define a score matrix 
${\bf D}\in\mathbb{R}^{K\times N}$, where $[{\bf D}]_{k,n}$ represents the 
confidence score of the token assigned to device $k$ in time slot $n$.

The estimated tokens whose confidence scores are smaller than a prescribed 
threshold are treated as masked tokens. In this subproblem, the receiver 
leverages the semantic context of the whole token sequence to predict these 
masked tokens. The detailed construction of the score matrix, the candidate 
token set, and the transformer-based masked token prediction procedure will be 
presented in Section~\ref{Sec:TokAssignPre}. {The reliability of such 
context-aware masked token prediction will be further quantified by the semantic 
orthogonality metric introduced in Section~\ref{Sec:SO}.
}

\section{Proposed CS-based Active Token Detection Scheme}
\label{Sec:TokDet}
To solve the high-dimensional underdetermined token detection problem (\ref{subprob1}), CS techniques offer an efficient solution, achieving accurate detection in scenarios where $L \ll Q$. In this section, we propose an approximate message passing (AMP)-based detection algorithm. In the following part, we describe the proposed detection algorithm for a given time slot $n\in[N]$, and hence, we drop the subscript $n$ for notational simplicity.

\subsection{Preliminaries}
Define $h_{q,m} = \left[{\bf H}\right]_{q,m}$, $y_{l,m} = \left[{\bf Y}\right]_{l,m}$, and $u_{l,q} = \left[{\bf U}\right]_{l,q}$.
The minimum mean square error (MMSE) estimate of ${\bf H}$ in (\ref{subprob1}) is the posterior mean, which can be expressed as
\begin{equation}\label{Eq:Post_Mean_Int}
{\hat h_{q,m}} = \int {h_{q,m}}p\left(h_{q,m}|{\bf Y} \right)d{h_{q,m}},\; \forall q,m.
\end{equation}
The marginal posterior distribution is given by
\begin{equation}\label{Eq:Mar_Post}
p\left(h_{q,m}|{\bf Y}\right) = \int p\left({\bf H}|{\bf Y}\right)d{\bf H}_{\backslash q,m},
\end{equation}
where ${\bf H}_{\backslash q,m}$ denotes the collection of the elements in ${\bf H}$ after removing $h_{q,m}$. The joint posterior distribution $p\left({\bf H}|{\bf Y}\right)$ can be computed according to the Bayesian rule as
\begin{equation}\label{Eq:Joint_Post}
\begin{aligned}
p\left({\bf H}|{\bf Y}\right) & = \frac{p\left({\bf Y}|{\bf H}\right)p_0\left({\bf H}\right)}{p\left({\bf Y}\right)} \\
                              & = \frac{1}{p\left({\bf Y}\right)}\prod\limits_{m = 1}^M{\left[\prod\limits_{l = 1}^L{p\left(y_{l,m}|{\bf H}\right)}\prod\limits_{q = 1}^Q{p_0\left(h_{q,m}\right)}\right]},
\end{aligned}
\end{equation}
where $p\left({\bf Y}\right)$ is a normalization factor, and the prior distribution and the likelihood function are respectively given by
\begin{equation}\label{Eq:Spike_Slab}
\begin{aligned}
\!\!\!p_0\!\left({\bf H}\right) \!\!=\!\!\! \prod\limits_{m = 1}^M{\prod\limits_{q = 1}^Q{\left[\left(1 - \gamma_{q}\right)\delta\left(h_{q,m}\right)+ {\gamma_{q}}\mathcal{CN}\left(h_{q,m};0,1\right)\right]}},\!\!\!
\end{aligned}
\end{equation}
\begin{equation}\label{Eq:Likelihood}
p\left(y_{l,m}|{\bf H}\right) \! \! = \! \! \frac{1}{\pi\sigma^2}\exp\left( \! - \frac{1}{\sigma^2}\left|y_{l,m}  \!-  \!\sum\limits_q {u_{l,q}}{h_{q,m}}\right|^2\right).
\end{equation}
In (\ref{Eq:Spike_Slab}), $0 \leq \gamma_{q} \leq 1$ is the sparsity ratio, i.e., the probability of $h_{q,m}$ ($m\in[M]$) being non-zero, $\delta\left(\cdot\right)$ is the Dirac delta function, and the distribution of the non-zero entries is due to Rayleigh fading channel assumption in (\ref{MA-SignalMod}).

\subsection{AMP-Based Posterior Mean Estimation}
As calculating the marginal posterior probabilities in (\ref{Eq:Mar_Post}) is difficult, especially for large signal dimension $Q$, we employ the AMP algorithm to realize an approximate posterior mean estimation of ${\bf H}$ with low complexity.

\begin{prop}
According to \cite{donoho2010message}, in the large system limit, i.e., as $Q\to \infty$ while $\frac{K}{Q}$ and $\frac{L}{Q}$ are fixed, the AMP algorithm decouples the matrix estimation problem in (\ref{subprob1}) into $QM$ scalar estimation problems. The posterior distributions of $h_{q,m}$, $\forall q,m$, are approximated as
\begin{equation}\label{Eq:Post_Approx1}
\begin{aligned}
p\left(h_{q,m}|{\bf Y}\right) &\approx p\left(h_{q,m}|R_{q,m}^t, \Sigma_{q,m}^t\right) \\
                              &\approx \frac{1}{\widetilde Z}p_0\left(h_{q,m}\right){\cal CN}\left(h_{q,m}; R_{q,m}^t, \Sigma_{q,m}^t\right),
\end{aligned}
\end{equation}
where $t$ denotes the $t$-th AMP iteration and ${\widetilde Z}$ is a normalization factor.
Messages $\Sigma_{q,m}^t$ and $R_{q,m}^t$ in (\ref{Eq:Post_Approx1}) are updated as
\begin{align}
\Sigma_{q,m}^t &= \left[\sum\nolimits_l \frac{\left|u_{l,q}\right|^2}{\sigma^2 + V_{l,m}^t}\right]^{-1}, \label{Eq:Var_Update1} \\
R_{q,m}^t &= {\hat h_{q,m}^t} + \Sigma_{q,m}^t\sum\nolimits_l{\frac{u_{l,m}^*\left(y_{l,m} - Z_{l,m}^t\right)}{\sigma^2 + V_{l,m}^t}}, \label{Eq:Var_Update2}
\end{align}
where messages $V_{l,m}^t $ and $Z_{l,m}^t$ are updated as
\begin{align}
V_{l,m}^t &= \sum\nolimits_q{\left|u_{l,q}\right|^2{v_{q,m}^t}}, \label{Eq:Fac_Update1}\\
Z_{l,m}^t &= \sum\nolimits_q{u_{l,q}{\hat h_{q,m}^t} - \frac{V_{l,m}^t}{\sigma^2 + V_{l,m}^{t - 1}}\left(y_{l,m} - Z_{l,m}^{t - 1}\right)}. \label{Eq:Fac_Update2}
\end{align}
\end{prop}
For detailed derivations, the readers are referred to \cite{donoho2010message}.
Furthermore, by applying (\ref{Eq:Spike_Slab}) in (\ref{Eq:Post_Approx1}), the posterior distribution can be reformulated as
\begin{equation}\label{Eq:Post_Approx2}
\begin{aligned}
\!\!\! p\left(h_{q,m}|R_{q,m}^t, \Sigma_{q,m}^t\right) &= \left(1 - \pi_{q,m}^t\right)\delta\left(h_{q,m}\right) \\
& + \pi_{q,m}^t{\cal CN}\left(h_{q,m}; {\mu}_{q,m}^t, {\tau}_{q,m}^t\right), \!\!\!
\end{aligned}
\end{equation}
where
\begin{align}
{\mu}_{q,m}^t &= \frac{{R_{q,m}^t}}{1 + \Sigma_{q,m}^t},\; {\tau}_{q,m}^t = \frac{{\Sigma_{q,m}^t}}{1 + \Sigma_{q,m}^t}, \label{Eq:Var_Mean_Var} \\
\pi_{q,m}^t &= \frac{\gamma_{q}^{t-1}}{\gamma_{q}^{t-1} + \left(1 - \gamma_{q}^{t-1}\right)\exp\left( - {\cal L}\right)}, \label{Eq:Belief_Indicator} \\
{\cal L} &= \ln{\frac{\Sigma_{q,m}^t}{1 + \Sigma_{q,m}^t}} + \left|R_{q,m}^t\right|^2 \left(\frac{1}{\Sigma_{q,m}^t} - \frac{1}{1 + \Sigma_{q,m}^t}\right), \label{Eq:L}
\end{align}
and $\pi_{q,m}^t$ is referred to as the activity indicator. Accordingly, the posterior mean $\hat h_{q,m}^t$ and variance $v_{q,m}^t$ can now be explicitly calculated as
\begin{align}
\hat h_{q,m}^t &= {\pi_{q,m}^t}{\mu_{q,m}^t}, \label{Eq:Post_Mean} \\
v_{q,m}^t &= {\pi_{q,m}^t}\left(\left|\mu_{q,m}^t\right|^2 + \tau_{q,m}^t\right) - \left|\hat h_{q,m}^t\right|^2, \label{Eq:Post_Var}
\end{align}
respectively. Therefore, for any $n\in[N]$, by iteratively executing (\ref{Eq:Var_Update1})-(\ref{Eq:Fac_Update2}), (\ref{Eq:Post_Mean}), and (\ref{Eq:Post_Var}) for $T_0$ iterations until convergence, a near-optimal MMSE estimate of ${\bf H}_n$, denoted by $\widehat{\bf H}_n$, can be obtained.

\subsection{Parameter Estimation and Token Detection}

Note that, in the prior distribution (\ref{Eq:Spike_Slab}), the sparsity ratio
$\gamma_q$, $\forall q\in[Q]$, is unknown. {Before iterative updation, we initialize
the sparsity ratio according to the phase-transition
curve based on the state evolution (SE) used in \cite{vila2013expectation}. Specifically, the initial sparsity ratio is set
as 
\begin{equation}\label{Eq:GammaInitSE}
\gamma_q^0
=
\frac{L}{Q}\rho_{\rm SE}\left(\frac{L}{Q}\right),
\quad \forall q\in[Q],
\end{equation}
where
\begin{equation}\label{Eq:RhoSE} \rho_{\rm SE}\left(\frac{L}{Q}\right) = \max_{c>0} \frac{ 1-\frac{2Q}{L} \left[ (1+c^2)\Phi(-c)-c\varphi(c) \right] }{ 1+c^2 - 2\left[ (1+c^2)\Phi(-c)-c\varphi(c) \right] }. \end{equation}
Here, $\Phi(\cdot)$ and $\varphi(\cdot)$ denote the cumulative distribution
function and probability density function of the standard normal distribution,
respectively. This initialization depends only on the measurement ratio $L/Q$
and does not require prior knowledge of the number of active devices.}

In each iteration, $\gamma_q^t$ is updated using the expectation maximization
(EM) algorithm \cite{moon1996expectation}. The EM update for $\gamma_q$ can be
expressed as
\begin{align}
\gamma_q^{t}
=
\arg\mathop{\max}\limits_{\gamma_q}
{\mathbb E}
\left[
\ln p\left({\bf H},{\bf Y}\right)
\mid
{\bf Y};\gamma_q^{t-1}
\right],
\label{Eq:EM}
\end{align}
where ${\mathbb E}[\cdot\mid{\bf Y};\gamma_q^{t-1}]$ denotes the expectation
conditioned on the received signal ${\bf Y}$ with parameter
$\gamma_q^{t-1}$. Thanks to the AMP decoupling,
$p({\bf H},{\bf Y})$ can be calculated based on
(\ref{Eq:Post_Approx2}). Hence, the update rule is obtained as
\begin{align}
\gamma_q^{t}
=
\frac{1}{M}\sum_{m=1}^M\pi_{q,m}^t,
\label{Eq:EMupdate}
\end{align}
where the estimation accuracy of $\gamma_q$ improves with the number of
observations $M$.

Furthermore, using threshold $T_h^r$, $0< T_h^r < 1$, the estimated {\it active token set} $\widehat{\mathcal{P}}_n$ is obtained as
\begin{equation}\label{eq:tokenDetection}
\left\{
\begin{aligned}
    &q\in \widehat{\mathcal{P}}_n, \quad \text{if} \quad (\gamma_q^{T_0})^n > T_h^r, \\
    &q \notin \widehat{\mathcal{P}}_n, \quad \text{if} \quad (\gamma_q^{T_0})^n\leq T_h^r,
\end{aligned}
\right.
\end{equation}
where $(\cdot)^n$ denotes the $n$-th time slot, $\forall n\in[N]$. Then, the set of estimated CSI $\widehat{\mathcal{F}}$ can be obtained as 
$\widehat{\mathcal{F}} = \{{\bf h}_{\phi, n}| {\bf h}_{\phi, n} = [\widehat{\bf H}_n]_{\phi, :}, \phi\in \widehat{\mathcal{P}}_n, n\in[N]\}$. 

{
The subsequent clustering-based token assignment requires the number of active
devices as the number of clusters. Since this quantity is generally unknown in
grant-free access, we estimate the instantaneous number of active devices from
the received signal and denote it by $\widehat K$. Motivated by the
rank-selection idea used in random access problems
\cite{shao2019dimension,ke2023next}, we estimate $\widehat K$ from the
dominant singular values of a stacked received signal matrix.

Specifically, we stack $N_s$ received token positions, indexed without loss of
generality by $1,\ldots,N_s$, as $\overline{\bf Y}
=
[{\bf Y}_1;\ldots;{\bf Y}_{N_s}]
\in\mathbb C^{N_sL\times M}$,
where the semicolon denotes vertical concatenation. Based on
(\ref{MA-SignalMod}), the stacked received signal can be expressed as
\begin{align}
    \overline{\bf Y}
=
\sum\nolimits_{k\in[K]}\overline{\bf x}_k({\bf h}_k)^T
+
\overline{\bf Z},
\end{align}
where
$\overline{\bf x}_k=[{\bf x}_{k,1};\ldots;{\bf x}_{k,N_s}]
\in\mathbb C^{N_sL}$ is the vertically stacked transmitted signal of device
$k$ over the selected $N_s$ token positions, and $\overline{\bf Z}$ denotes the
corresponding stacked noise matrix. In the noiseless case, we have
$\overline{\bf Y}_{0}
=
\sum_{k\in[K]}\overline{\bf x}_k({\bf h}_k)^T$, which is a sum of $K$ rank-one
components. If $N_sL\ge K$, $M\ge K$, and both
$\{\overline{\bf x}_k\}_{k=1}^K$ and $\{{\bf h}_k\}_{k=1}^K$ are linearly
independent, then ${\rm rank}(\overline{\bf Y}_{0})=K$. Therefore, the number
of active devices can be inferred from the signal subspace of the noisy stacked
received matrix.

Let $\lambda_1\ge \lambda_2\ge\cdots$ denote the singular values of
$\overline{\bf Y}$. We estimate the number of active devices as
\begin{equation}\label{Eq:Khat}
\widehat K
=
\arg\max_{1\le r\le r_{\max}}
\frac{\lambda_r}{\lambda_{r+1}+\epsilon},
\end{equation}
where $r_{\max}<\min(N_sL,M)$ is the maximum candidate number of active devices,
and $\epsilon>0$ is a small constant for numerical stability. The estimated
$\widehat K$ is used as the number of clusters in the subsequent token
assignment stage.

}


\section{Proposed Token Assignment and Context-Aware Masked Token Prediction}
\label{Sec:TokAssignPre}

Under the slow-fading channel assumption, the channel vector of each active device is approximately invariant over the $N$ time slots. Hence, the estimated CSI in $\widehat{\mathcal F}$ are expected to form device-dependent clusters. We therefore begin with a coarse token assignment by clustering $\widehat{\mathcal F}$ into $\widehat K$ groups, where $\widehat K$ denotes the estimated number of active devices. This step associates the detected active tokens across different time slots with $\widehat K$ estimated device streams. In the absence of token collisions and severe detection errors, such clustering can effectively resolve the token assignment problem in~(\ref{subprob2-2}). However, when token collisions or unreliable detections occur, some token positions may remain ambiguous or unfilled in the estimated token sequences, which are referred to as \emph{masked positions}. To address this issue, after coarse assignment, we construct a score matrix to refine the assignment and identify a \emph{candidate token set} for the masked positions. The masked tokens are then predicted by exploiting both the semantic context and the candidate token set.


\subsection{Clustering-Based Coarse Token Assignment}
Given the estimated active token sets $\{\widehat{\mathcal{P}}_n\}_{n=1}^{N}$, 
the estimated CSI set $\widehat{\mathcal{F}}$, and the 
estimated number of active devices $\widehat K$, we first group the channel 
vectors in $\widehat{\mathcal{F}}$ into $\widehat K$ clusters. The clustering problem can be formulated as follows
\begin{align}\label{eq:clustering}
    &\min_{\{\mathcal{C}_1, \mathcal{C}_2, \dots, \mathcal{C}_{\widehat K}\}} 
    \sum_{{k}=1}^{\widehat K} \sum_{\mathbf{h} \in \mathcal{C}_{{k}}} 
    \|\mathbf{h} - \mathbf{c}_{{k}}\|_2^2,\nonumber \\
    \text{s.t.} & \quad 
    \bigcup_{{k}=1}^{\widehat K} \mathcal{C}_{{k}} = \widehat{\mathcal{F}}, \quad
    \mathcal{C}_i \cap \mathcal{C}_j = \varnothing, \quad i \neq j.
\end{align}
where $\mathbf{c}_k = \frac{1}{|\mathcal{C}_k|} \sum_{\mathbf{h} \in \mathcal{C}_k} \mathbf{h}$, and the partition $\{\mathcal{C}_1, \mathcal{C}_2, \dots, \mathcal{C}_{\widehat{K}}\}$ is found by minimizing the sum of squared distances between each vector and its cluster centroid. We resort to the Kmeans++ algorithm \cite{arthur2006k} to determine the $\widehat{K}$ cluster centers $\mathbf{c}_k$, as well as the partition $\{\mathcal{C}_1, \mathcal{C}_2, \dots, \mathcal{C}_{\widehat K} \}$. For notational simplicity, we use $k\in[\widehat K]$ to index the estimated devices in this section.

Then, the tokens can be assigned as follows
\begin{align}\label{eq:coarseAssign}
    \left[ \widehat{\bf B}_k\right]_{\phi, n} = 1,~~\text{if}~~{\bf h}_{\phi, n}\in\mathcal{C}_k,~\text{where}~\phi\in \widehat{\mathcal{P}}_n,~\forall n,k.
\end{align}

\subsection{Fine-Grained Token Assignment}

\subsubsection{Score Matrix Definition}

After the coarse assignment, we introduce a score matrix
${\bf D}\in\mathbb{R}^{\widehat K\times N}$ to quantify the confidence
of the token-to-cluster assignment. Specifically, for $k\in[\widehat K]$
and $n\in[N]$, the elements of ${\bf D}$ are defined as
\begin{equation}\label{eq:score_matrix}
\left[{\bf D}\right]_{k,n}
=
\begin{cases}
\dfrac{1}{\left\|{\bf h}_{\phi,n}-{\bf c}_k\right\|_2},
&
\text{if }
\left[\widehat{\bf B}_k\right]_{\phi,n}=1,\;
\phi\in\widehat{\mathcal P}_n,
\\[2mm]
0,
&
\text{if }
\left[\widehat{\bf B}_k\right]_{:,n}
=
{\bf 0}_{Q\times 1}.
\end{cases}
\end{equation}
A larger value of $\left[{\bf D}\right]_{k,n}$ indicates that the
estimated channel vector associated with the assigned token is closer to
the corresponding cluster center, and hence the assignment is more
reliable. In contrast, $\left[\widehat{\bf B}_k\right]_{:,n} = {\bf 0}_{Q \times 1}$ means that no token is assigned to device $k$ in time slot $n$ in the first stage; hence, the score is zero.

\subsubsection{Low-Confidence Assignment Identification}

Based on the score matrix and a prescribed threshold $T_h^s$ on the assignment confidence scores, we identify the set of low-confidence assigned tokens in the $n$-th time slot as
\begin{equation}\label{eq:IncorrectTokens} \widetilde{\mathcal P}_n = \left\{ \phi \mid \left[\widehat{\bf B}_k\right]_{\phi,n}=1,\, 0<\left[{\bf D}\right]_{k,n}<T_h^s \right\}.
\end{equation}
We set $T_h^s=1/(\frac{2}{|\widehat{\mathcal{F}}|}\sum_{{k}=1}^{\widehat K} \sum_{\mathbf{h} \in \mathcal{C}_k} \|\mathbf{h} - \mathbf{c}_k\|_2)$, which is the reciprocal of twice the average Euclidean distance between all CSI and their respective cluster centers. 

\begin{figure}[t]
    \centering
    \captionsetup{font={footnotesize, color={black}}, singlelinecheck=off, justification=raggedright, name={Fig.}, labelsep=period}
    
    \captionsetup[subfloat]{justification=centering}
    
    \subfloat[Token assignment in a given time slot.]{
        \includegraphics[width=0.76\columnwidth, keepaspectratio]{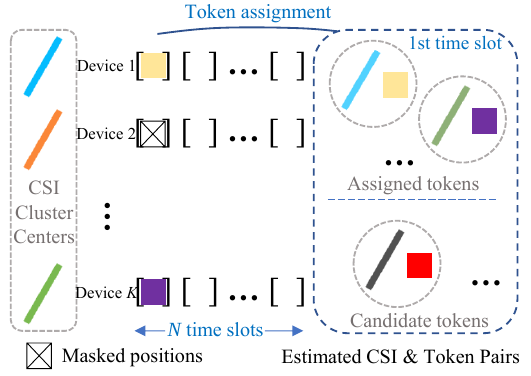}
        \label{figTokMatch1}
    }\\
    \subfloat[Masked token predictions of a given device.]{
        \includegraphics[width=0.76\columnwidth, keepaspectratio]{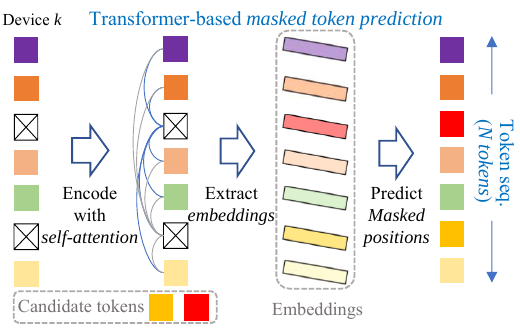} 
        \label{figTokMatch2}
    }
    \caption{Illustration of token assignment and masked token predictions.}
    \label{figTokMatch}
\end{figure}

The low-confidence tokens in $\widetilde{\mathcal P}_n$ are retained as candidate tokens for the subsequent masked token prediction stage. Therefore, $\widetilde{\mathcal P}_n$ is referred to as the \emph{candidate token set}. For each low-confidence assignment, we mask the corresponding position as \begin{equation}\label{eq:BkUpdate} \left[\widehat{\bf B}_{k}\right]_{:,n} = {\bf 0}_{Q\times 1}, \quad \text{if} \quad 0<\left[{\bf D}\right]_{k,n}<T_h^s . \end{equation}
Token
assignment, candidate token sets, and masked positions are illustrated in
Fig.~\ref{figTokMatch}.

\subsection{Transformer-Based Masked Token Prediction}
For each estimated token sequence, a position $(k,n)$ is regarded as a masked position if the corresponding column of $\widehat{\bf B}_k$ is an all-zero vector, i.e., $\left[\widehat{\bf B}_k\right]_{:,n} = {\bf 0}_{Q\times 1}$. Such a masked position may occur because no token is assigned during the coarse assignment stage or because a low-confidence assignment is removed in the fine-grained assignment stage. Before being fed into the transformer, the all-zero columns in $\widehat{\bf B}_k$ are replaced by the corresponding modality-specific mask tokens.

{
Let $h_{\boldsymbol\theta}(\cdot)$ denote the pre-trained transformer with parameters $\boldsymbol\theta$. Given the $k$-th masked token sequence, the transformer outputs a contextual probability distribution over the whole token codebook for each masked position
\begin{equation}\label{eq:transformer_prediction} 
{\bf P}_k = h_{\boldsymbol\theta}(\widehat{\bf B}_k), 
\end{equation} 
where ${\bf P}_k\in\mathbb{R}^{Q\times N}$ and $\sum_{q=1}^{Q}[{\bf P}_k]_{q,n}=1$, $\forall k,n$. 
The element $[{\bf P}_k]_{q,n}$ represents the probability that the masked position $(k,n)$ should be recovered as the $q$-th token in the codebook according to the semantic context of the entire token sequence. For each masked position $(k,n)$, the recovery rule depends on the cardinality of the candidate token set $\widetilde{\mathcal P}_n$. the recovered token index $q_{k,n}^{\star}$ and the corresponding one-hot assignment are given by
\begin{equation}\label{eq:unified_prediction} \left[\widehat{\bf B}_k\right]_{q^*,n} = 1, \quad q_{k,n}^{\star} = \begin{cases} \displaystyle \arg\max_{q\in[Q]} \left[{\bf P}_k\right]_{q,n}, & |\widetilde{\mathcal P}_n|=0, \\[3mm] \phi,~\phi\in\widetilde{\mathcal P}_n, & |\widetilde{\mathcal P}_n|=1,\; \\[1mm] \displaystyle \arg\max_{q\in\widetilde{\mathcal P}_n} \left[{\bf P}_k\right]_{q,n}, & |\widetilde{\mathcal P}_n|>1, \end{cases} \end{equation} 


The three cases in \eqref{eq:unified_prediction} correspond to different levels of physical-layer side information. If $|\widetilde{\mathcal P}_n|=0$, no candidate token is available from the physical-layer assignment stage, and the transformer predicts the masked token from the whole codebook. If $|\widetilde{\mathcal P}_n|=1$, the candidate token is unique, and the masked position can be directly recovered without semantic disambiguation. If $|\widetilde{\mathcal P}_n|>1$, the physical-layer observations narrow the possible tokens down to a candidate set, and the transformer selects the most contextually consistent token from this set. 
}


Equivalently, one may sample the recovered token from the predicted categorical distribution restricted to the corresponding search space. In this work, we adopt the deterministic maximum-a-posteriori rule for reproducible evaluation. The function $h_{\boldsymbol\theta}(\cdot)$ is difficult to construct analytically because it requires modeling long-range contextual dependencies among tokens. Fortunately, transformer-based generative models provide a powerful data-driven solution. A well-trained bidirectional transformer can infer the distribution of a masked token from its surrounding context and can therefore exploit semantic orthogonality among different token sequences. Here, we leverage pre-trained bidirectional transformer networks, namely BERT for text tokens and MaskGIT for image tokens, to perform masked token prediction\footnote{ We adopt masked generation rather than autoregressive generation because the receiver observes an incomplete token sequence with missing positions distributed over the entire block. Masked generation can exploit both preceding and succeeding contextual tokens and can predict multiple masked positions in parallel.}. The prediction process is illustrated in Fig.~\ref{figTokMatch2}. For simplicity, the transformer performs a single inference for each estimated device stream and predicts all remaining masked positions simultaneously. A multi-step prediction strategy, where only a subset of masked tokens is predicted at each step, may further improve the reconstruction performance and is left for future work. 

{
\begin{remark} The candidate-restricted recovery rule in \eqref{eq:unified_prediction} uses the physical-layer detection and assignment results as side information for masked token prediction. This is reasonable because $\widetilde{\mathcal P}_n$ is likely to contain the transmitted token when active token detection is accurate. As will be demonstrated in Section~\ref{Sec:Simul}, the token detection error decreases as the number of antennas at the BS increases, which is desirable for massive MIMO systems deployed in 5G and beyond wireless networks. In other words, the active token detection error can be made very small for large $M$ without increasing the communication overhead. 

The case $|\widetilde{\mathcal P}_n|=1$ can be interpreted as a
degenerate case of candidate-restricted masked token prediction, where
the search space contains only one token. This situation typically occurs
when the ambiguity is caused by multiple devices transmitting the same
token in the same time slot. In this case, direct recovery avoids
unnecessary full-codebook prediction. In low-SNR regimes, however, missed
detections and false alarms may lead to an inaccurate candidate set, and
the singleton update may become suboptimal. Nevertheless, replacing this
update by full-codebook prediction does not necessarily improve the
reconstruction performance, since the search space increases from one
candidate token to $Q$ possible tokens. The impact of imperfect
physical-layer detection is included in the end-to-end simulation results.
\end{remark}

\subsection{Semantic-Orthogonality Interpretation of Candidate-Restricted Masked Prediction}
\label{Sec:SO}

The key idea of ToDMA is that unresolved token ambiguities can be resolved not
only by physical-layer observations, but also by semantic context. Therefore,
semantic orthogonality is quantified over the unresolved candidate token set,
rather than over the whole token vocabulary.

For a masked position $(k,n)$ with $|\widetilde{\mathcal P}_n|>1$, the
transformer output ${\bf P}_k$ provides a contextual probability distribution
over the whole token codebook. We restrict this distribution to the candidate
token set $\widetilde{\mathcal P}_n$ as
\begin{align}
p_k(q,n)
=
\frac{[{\bf P}_k]_{q,n}}
{\sum_{q'\in\widetilde{\mathcal P}_n}[{\bf P}_k]_{q',n}},
\quad q\in\widetilde{\mathcal P}_n.
\end{align}
Then, the token-domain semantic orthogonality is defined as
\begin{align}
\xi_{k,n}
=
1-
\frac{
H\left(p_k(\cdot,n)\right)
}{
\log|\widetilde{\mathcal P}_n|
},
\end{align}
where
\begin{align}
H\left(p_k(\cdot,n)\right)
=
-\sum_{q\in\widetilde{\mathcal P}_n}
p_k(q,n)\log p_k(q,n).
\end{align}
Since $0\leq H(p_k(\cdot,n))\leq \log|\widetilde{\mathcal P}_n|$, we have $0\leq \xi_{k,n}\leq 1$. A larger $\xi_{k,n}$ indicates that the semantic context sharply favors one candidate token, which means that the candidate tokens are highly distinguishable under the current context. In contrast, a smaller $\xi_{k,n}$ implies that the restricted posterior is close to uniform, and the candidate tokens remain semantically difficult to distinguish. 

The indicator $\xi_{k,n}$ provides a statistical interpretation of the
effectiveness of candidate-restricted masked token prediction. Since the
transformer output reflects a learned contextual distribution over the token
codebook, induced by the global source statistics captured during training, a
larger $\xi_{k,n}$ indicates that the semantic context provides a stronger
preference for one candidate token. Therefore, $\xi_{k,n}$ is expected to
correlate with a higher recovery probability on average, which can be verified
through Monte-Carlo simulations. Nevertheless, $\xi_{k,n}$ should not be
interpreted as a strict correctness certificate for each individual
realization, since the predicted distribution may still assign the largest
probability to an incorrect token.

This interpretation also highlights the complementary roles of the physical layer and the semantic model in ToDMA. The physical-layer receiver narrows the search space from the whole codebook $[Q]$ to the candidate token set $\widetilde{\mathcal P}_n$, while the transformer uses the surrounding context to select the most plausible token within this reduced set. Therefore, even if multiple candidate tokens are physically ambiguous, they may still be separable in the semantic domain if their contextual probabilities are sufficiently different. 
}

{
\begin{remark}
\label{re:nonuniform}
In practice, devices may transmit messages with heterogeneous token
lengths, owing to variable-length text or images encoded at different
resolutions or quality levels. ToDMA supports this setting by inserting
beginning-of-message and end-of-message tokens. A device terminates its
transmission after sending the end-of-message token and therefore requires
no redundant padding. The effectiveness of this variable-length operation
is validated in Section~\ref{Sec:TextTransmission}.
\end{remark}
}

\section{Proposed ToDMA Receiver Algorithm}
In this section, we summarize the ToDMA receiver design and analyze its computational complexity.

\subsection{Algorithm Summary}
The proposed ToDMA receiver design is summarized in {\bf Algorithm \ref{Algo:Receiver}}. Note that the device identities are unknown; hence, the receiver returns an unordered set of $K$ estimated token sequences. 

\begin{algorithm}[t]
\small 
\caption{ToDMA Receiver Design}
\label{Algo:Receiver}
\SetAlgoLined
\KwIn{Received signal ${\bf Y}_n$, modulation codebook ${\bf U}$, noise variance $\sigma^2$.}
\KwOut{Estimated token sequences \{$\widehat{\bf B}_k$, $\forall k\in[K]$\}.}

\tcp{Token detection}

\textbf{Initialization:} $\forall n,q,m,l$, $(\hat h_{q,m}^1)^n=0$, $(v_{q,m}^1)^n=1$, $(Z_{l,m}^0)^n=y_{l,m}$, $(\gamma_q^0)^n=0.5$\;
\label{Alg1:ini_AMP}
\textbf{Iteration:} For $t=1,...,T_0$, execute (\ref{Eq:Var_Update1})-(\ref{Eq:Fac_Update2}), (\ref{Eq:Post_Mean}), (\ref{Eq:Post_Var}), and (\ref{Eq:EMupdate})\;
\label{Alg1:AMP_ite}
\textbf{Detection:} {Find $\widehat{\mathcal{P}}_n$, $\widehat{\mathcal{F}}$, and $\widehat{K}$ according to (\ref{eq:tokenDetection}) and (\ref{Eq:Khat}).} 
\label{Alg1:AMP_output}

\tcp{Token assignment}
\textbf{Initialization:} $\forall k$, $\widehat{\bf B}_k={\bf 0}_{Q\times N}$, \;
\label{Alg1:Assign_ini}
\textbf{Coarse assignment:} Using Kmeans++ to group $\widehat{\mathcal{F}}$ into $\widehat{K}$ clusters, then execute (\ref{eq:coarseAssign}) according to $\widehat{\mathcal{P}}_n$\;
\label{Alg1:Assign_1}
\textbf{Fine-grained assignment:} Given $\widehat{\mathcal{P}}_n$, execute (\ref{eq:score_matrix}), (\ref{eq:IncorrectTokens}), and (\ref{eq:BkUpdate}) in order\;
\label{Alg1:Assign_2}
\tcp{Masked token predictions}
\textbf{Finalize token assignment:} Execute (\ref{eq:unified_prediction})\;
\label{Alg1:MTP}
\textbf{Return:} \{$\widehat{\bf B}_k$, $\forall k\in[K]$\}\;
\label{Alg1:Output}
\end{algorithm}

\subsection{Computational Complexity}
The computational complexity of Algorithm \ref{Algo:Receiver} is mainly caused by three components.
\begin{itemize}
    \item The AMP-based token detection, i.e., line \ref{Alg1:AMP_ite}, whose complexity is $\mathcal{O}\left( T_0 N L QM\right)$. It is linear with respect to the signal dimensions $L$, $Q$, $M$, and the number of tokens $N$, and the number of iterations $T_0$.
    \item The clustering algorithm used in token assignment, i.e., line \ref{Alg1:Assign_1}, whose complexity is $\mathcal{O}\left( T_c KNM\right)$, where $T_c$ denotes the corresponding number of iterations.
    \item {Transformer-based masked-token prediction in line~\ref{Alg1:MTP}, which computes the token probability distribution ${\bf P}_k$ in Section~VI-C. Taking BERT-Base as an example, its dominant computational complexity is $\mathcal{O}(KN^2H)$, where $H$ denotes the hidden dimension. The complete masked-prediction cost for ${\bf P}_k$, together with the per-device tokenizer and detokenizer costs omitted from the asymptotic analysis, is reported in Giga-floating-point operations (GFLOPs) in Table~\ref{tab:complexity}.}
\end{itemize}

Hence, the overall computational complexity can be calculated as $\mathcal{O}\left[(T_0 N L QM + K\left( T_cNM + N^2 H\right) \right]$. Hence, the large tokenizer size $Q$ is decoupled from the number of devices $K$, which is desirable for serving many devices. In addition, the overall computational complexity is linearly proportional to the signal dimensions, except for the transformer part. Lightweight prediction models may further reduce the receiver-side computational overhead.



\section{Simulation Results}
\label{Sec:Simul}

In this section, we present the results of our extensive simulations to evaluate the performance of the proposed ToDMA framework.

\subsection{Parameter Settings}

The proposed ToDMA framework is applicable to various data modalities. Here, we will present results for image and text signals as examples. We first introduce the general simulation settings. Specific parameters related to different source modalities will be described later in the corresponding subsections.

\subsubsection{Wireless Parameters}
We consider $K_{\rm T} = 400$ devices, out of which $K$ active devices are randomly chosen, with $K$ taking values from $\{20, 40, 60, 80\}$. The number of receive antennas at the BS is $M = 256$. {For the activity-level estimator in (\ref{Eq:Khat}), we stack $N_s=16$ received token positions.} Unless otherwise specified, these values are used as default settings. The channel vectors ${\bf h}_k$, $\forall k \in [K]$, are generated according to the Rayleigh distribution. The elements of the communication codebook ${\bf U} \in \mathbb{C}^{L \times Q}$ are sampled from an i.i.d. complex Gaussian distribution, where $Q$ is the dimension of the token codebook.

\subsubsection{Token Communication Metrics}
The following general metrics for ToDMA are applicable to source signals of any modality:
\begin{itemize}
    \item {\bf NMSE}: The normalized mean squared error is defined as
    \begin{equation}\label{eq-nmse}
\text{NMSE} = 10\log_{10}\left( \dfrac{1}{N}\sum_{n=1}^N\|\widehat{\bf H}_n-{\bf H}_n\|_F/\|{\bf H}_n \|_F \right).
\end{equation}
    \item {\bf TDER}: The token detection error rate (TDER) is defined as
       \begin{equation}\label{eq-tder}
\text{TDER} = \dfrac{1}{N K}\sum_{n=1}^N \left|\left(\mathcal{P}_n \setminus \widehat{\mathcal{P}}_n \right) \cup \left(\widehat{\mathcal{P}}_n \setminus {\mathcal{P}}_n \right)\right|,
\end{equation}
where $\mathcal{P}_n \setminus \widehat{\mathcal{P}}_n$ represents the set of elements in $\mathcal{P}_n$ that do not belong to $\widehat{\mathcal{P}}_n$.
    \item {\bf TER}: The token error rate (TER) is defined as
       \begin{equation}\label{eq-ter}
\text{TER} = \dfrac{1}{2NK}\sum_{k=1}^{K}\|\widehat{\bf B}_k - {\bf B}_k\|_0.
\end{equation}
\end{itemize}
The proposed ToDMA does not rely on any prior knowledge of device identities during transmission or decoding. Device identity information is introduced only for performance evaluation when computing NMSE and TER, and thus, does not compromise the fairness of the comparison.

\begin{figure}[t]
\centering
\subfloat[TDER versus $L$]{
\includegraphics[width=0.43\columnwidth]{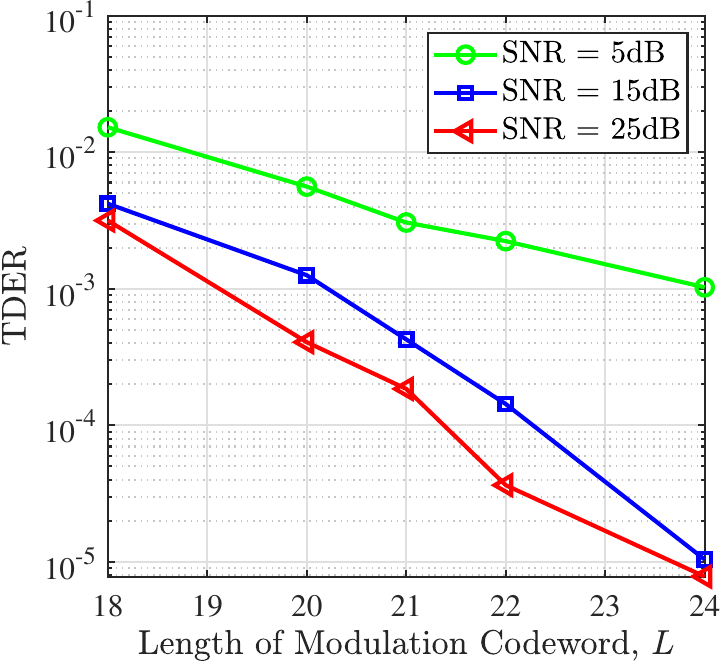}
    \label{FigTAa}
}
\subfloat[NMSE versus $L$]{    \includegraphics[width=0.43\columnwidth]{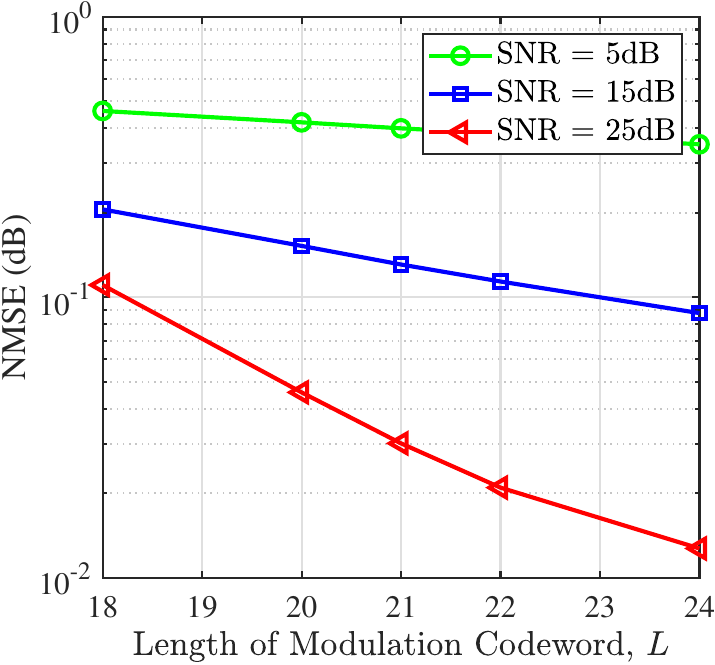}
    \label{FigTAb}
}
\captionsetup{font={footnotesize}, singlelinecheck = off, justification = justified,name={Fig. },labelsep=period}
\caption{Token detection performance versus length of modulation codeword $L$, with parameters set to $K=20$, $M=256$, and $Q=1024$.}
\label{Sim:Lchange}
\end{figure}

\begin{figure}[t]
\centering
\subfloat[TDER versus $M$]{
    \includegraphics[width=0.43\columnwidth]{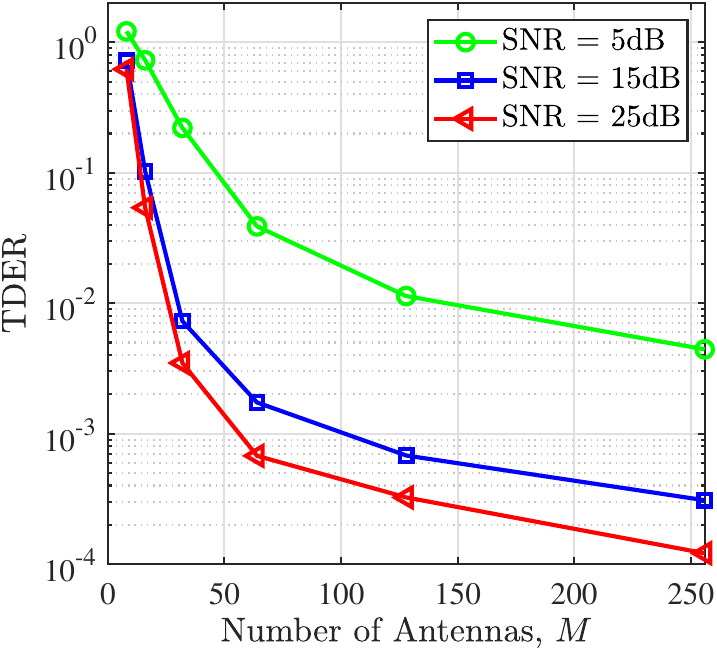}
    \label{FigTAa2}
}
\subfloat[NMSE versus $M$]{
    \includegraphics[width=0.43\columnwidth]{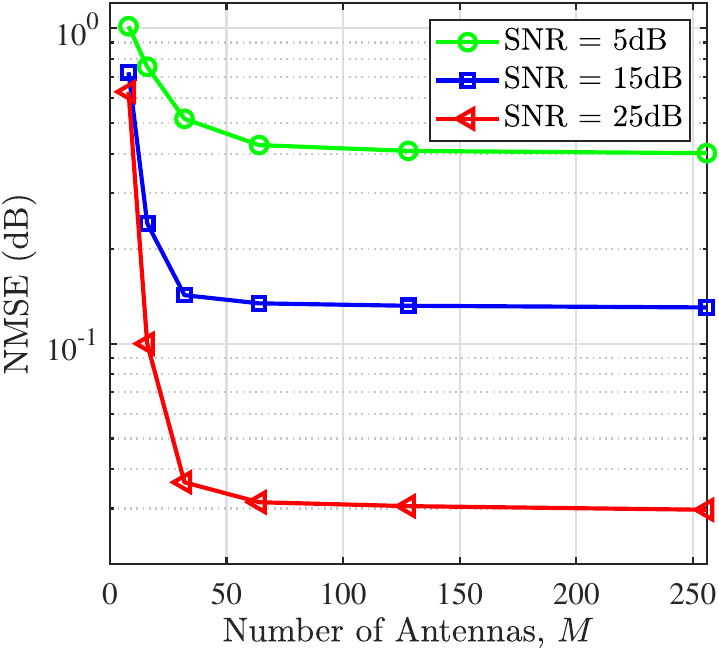}
    \label{FigTAb2}
}
\captionsetup{font={footnotesize}, singlelinecheck = off, justification = justified,name={Fig. },labelsep=period}
\caption{Token detection performance versus the number of antennas $M$, with parameters set to $K=20$, $L=K+1$, and $Q=1024$.}
\label{Sim:Mchange}
\end{figure}

{For a fixed token-alphabet size of $Q=1024$, Fig. \ref{Sim:Lchange} shows
that increasing $L$ provides more CS observations and significantly
improves TDER and NMSE across all signal-to-noise ratio (SNR) levels, at the cost of additional channel
uses or bandwidth. In Fig.~\ref{Sim:Mchange}, we fix $L=K+1$ and evaluate TDER and NMSE as functions of the number of BS antennas $M$. As $M$ increases, NMSE saturates at a noise-related floor, whereas TDER continues to decrease because multiple antennas provide multiple-measurement-vector rank diversity~\cite{davies2012rank}. Hence, for sufficiently large $M$, reliable token-set estimation can be achieved without increasing the communication overhead $L$. In practice, $L$ is selected according to the token-alphabet size, available bandwidth, and traffic conditions.}

Since TER is dependent on the data distribution of multiple devices, we will evaluate the TER performance on both image and text datasets in the following subsections.

\subsection{ToDMA-Based Image Transmission}
\subsubsection{Settings}
{\bf Dataset:} We consider the well-known ImageNet-100 dataset \cite{russakovsky2015imagenet}, consisting of 130000 images with resolutions $h\times w=256\times 256$. In each Monte-Carlo simulation, each active device randomly chooses one image from the dataset.
{\bf Tokenizer:} We employ the pre-trained VQGAN-based tokenizer from \cite{esser2021taming}. {This model is trained with a perceptual loss and an adversarial loss to ensure high visual fidelity in the discrete latent space.} The tokenizer has a codebook size of $Q=
1024$, with each image encoded as a sequence of $N=256$ tokens. 
{\bf Masked image prediction:} We employed the pre-trained vision transformer model described in \cite{chang2022maskgit}.

\subsubsection{Baseline Schemes}

{\textbf{Pretrained DJSCC}: We further consider a pretrained-tokenizer-aided DJSCC baseline, referred to as \textit{Pretrained DJSCC}. It uses the same pretrained VQGAN encoder and decoder as ToDMA, but does not employ MaskGIT, label guidance, or any semantic recovery module. Each image is first encoded into a continuous VQGAN latent tensor of size $16\times16\times256$. A lightweight CNN-based JSCC encoder then compresses this latent tensor into a fixed number of complex channel symbols, e.g., $256$ or $512$ symbols, for AWGN transmission. At the receiver, the JSCC decoder reconstructs the latent tensor, which is further decoded into an image by the frozen VQGAN decoder. During training, the VQGAN backbone is fixed, and only the JSCC encoder and decoder are optimized. To align this continuous-latent baseline with the discrete VQGAN representation used by ToDMA, the training objective combines latent-domain reconstruction, image-domain reconstruction, perceptual similarity, and a codebook-consistency regularization term. This regularization encourages each reconstructed latent vector to stay close to its corresponding VQGAN codebook entry. In the final reconstruction stage, the recovered latent vectors are projected to their nearest codebook entries before VQGAN decoding. Thus, this scheme provides a controlled continuous-latent JSCC baseline based on the same pretrained visual representation as ToDMA.
}
For \textit{Pretrained DJSCC}, we adopt a grant-based random access (GBRA) protocol to support multiple active devices. 
Specifically, the $K$ active devices first complete the scheduling-request and uplink-grant handshake, and the granted uplink resources are then partitioned orthogonally and equally among the active devices. 
In addition, we assume perfect channel equalization, so that the resulting transmission is modeled as an AWGN channel.

\textbf{Tokenized URA}: 
To isolate the gain of the proposed context-aware collision mitigation from that of tokenization and unsourced random access (URA), we introduce a context-agnostic \textit{Tokenized URA} baseline. 
This baseline uses the same tokenizer, channel code structure, and detection procedure as ToDMA. 
However, token assignment at the receiver is performed solely based on the estimated CSI, without exploiting semantic correlations among consecutive tokens. 
{For any remaining masked position, if candidate tokens are available, the receiver randomly selects one token from the candidate token set; otherwise, it randomly selects one token from the whole codebook.}
Therefore, the performance gap between ToDMA and Tokenized URA quantifies the gain contributed by the proposed context-aware de-collision mechanism.

\textbf{Ideal VQGAN}: 
We also report an \textit{Ideal VQGAN} baseline, where the same VQGAN tokenizer as ToDMA is used but all discrete tokens are assumed to be transmitted without error. 
This result characterizes the reconstruction quality limited only by semantic compression, rather than by channel impairments or multiuser collisions.

{
Existing NOMA-based DJSCC studies generally consider a fixed and relatively small number of identified users~\cite{yilmaz2025learning,Wang2025SemanticNOMA}, whereas ToDMA targets massive grant-free unsourced access with sporadically active devices. Advanced URA collision-mitigation methods can be incorporated into the candidate-detection stage of both ToDMA and Tokenized URA~\cite{YongpengUMA,YongpengJSAC}; therefore, under the same detector and candidate sets, their performance gap mainly isolates the gain of context-aware token recovery.
}

\begin{table}[t]
\centering
\caption{CBRs of different image transmission schemes}
\label{tab:cbr}
\setlength{\tabcolsep}{3pt} 
\begin{tabular}{c|>{\centering\arraybackslash}p{1.2cm}|c|c|c|c}
\toprule
\textbf{Scheme} & \textbf{CBR} 
& \textbf{$L=2K$} & \textbf{$L=K+1$} 
& \textbf{$N_{\rm sym}{=}512$} & \textbf{$N_{\rm sym}{=}256$} \\
\midrule
\makecell{Proposed\\ ToDMA} &
$\frac{LN}{3Khw}$ 
& 0.0026 
& $\approx$ 0.0013 
& N/A & N/A\\[2mm]

\makecell{Tokenized\\ URA} &
$\frac{LN}{3Khw}$ 
& 0.0026 
& $\approx$ 0.0013 
& N/A & N/A\\[2mm]

{\makecell{Pretrained\\ DJSCC}} &
{$\frac{N_{\rm sym}}{3hw}$} 
& {N/A}
& {N/A}
& {0.0026} 
& {0.0013}
\\[1mm]

\bottomrule
\end{tabular}

\end{table}

\subsubsection{Performance Metrics}
\textbf{Channel Bandwidth Ratio (CBR):} 
CBR measures the average number of channel uses (symbols) per source dimension. For an RGB image of size $h\times w$, it is defined as $\mathrm{CBR}=N_{\rm sym}/(3hw)$, where $N_{\rm sym}$ denotes the number of complex channel symbols used to transmit one image \cite{bourtsoulatze2019deep}. Table~\ref{tab:cbr} summarizes the resulting CBRs of ToDMA and Pretrained DJSCC. For ToDMA and tokenized URA, the CBR depends on the length of modulation codeword $L$. For Pretrained DJSCC, the channel encoder produces a $2N_{\mathrm{sym}}$-dimensional real-valued latent representation. Every two real-valued entries are mapped to the in-phase and quadrature components of one complex channel symbol, resulting in $N_{\mathrm{sym}}$ transmitted complex symbols per image. Furthermore, we consider {\bf CLIP} (Contrastive Language–Image Pre-training) and {\bf LPIPS} (Learned Perceptual Image Patch Similarity), which are widely used in semantic communications as perceptual quality metrics \cite{beyondBits,Magzine2025Token}, capturing semantic and human-aligned fidelity. In addition, {\bf PSNR} (Peak Signal-to-Noise Ratio) is considered as a conventional pixel-level distortion measure.

\begin{figure*}[!t]
    \centering
    \captionsetup{font={footnotesize}, labelsep=period} 
    \subfloat[]{
        \includegraphics[width=0.24\textwidth]{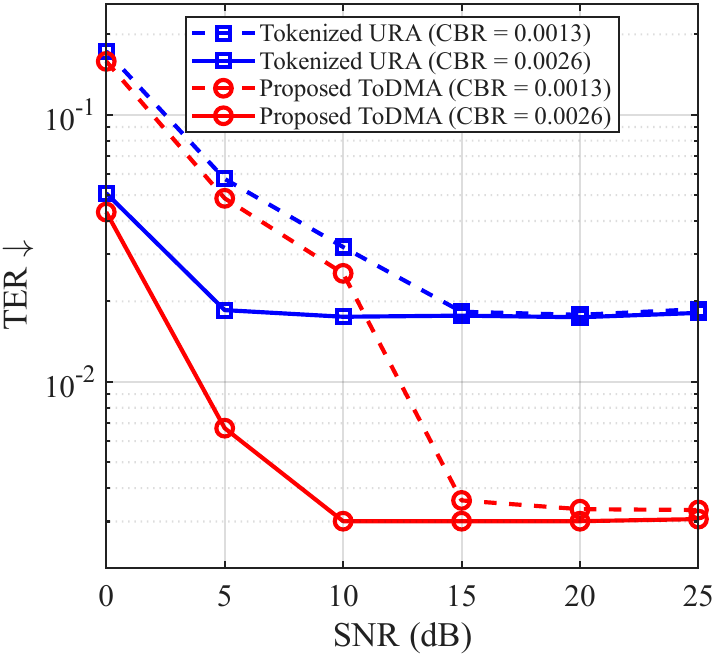}
        \label{fig:snr-a}
    }
    \subfloat[]{
        \includegraphics[width=0.24\textwidth]{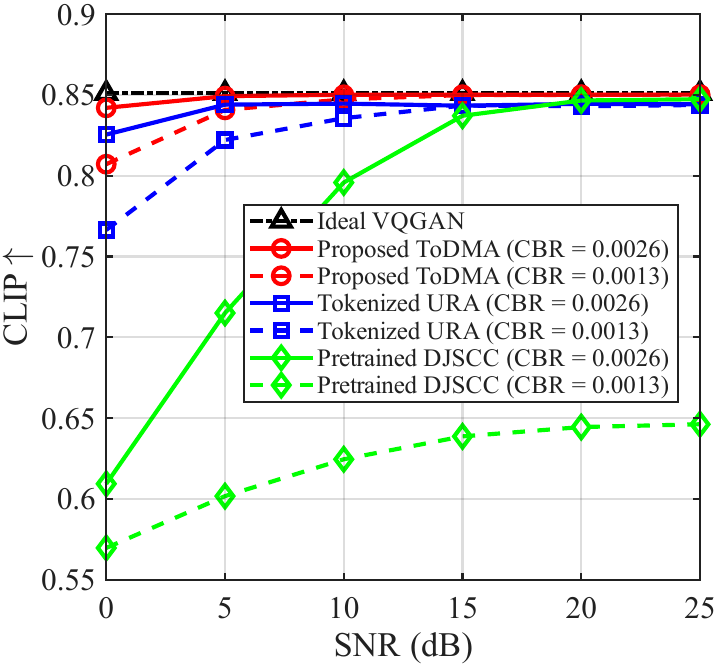}
        \label{fig:snr-b}
    }
    \subfloat[]{
        \includegraphics[width=0.24\textwidth]{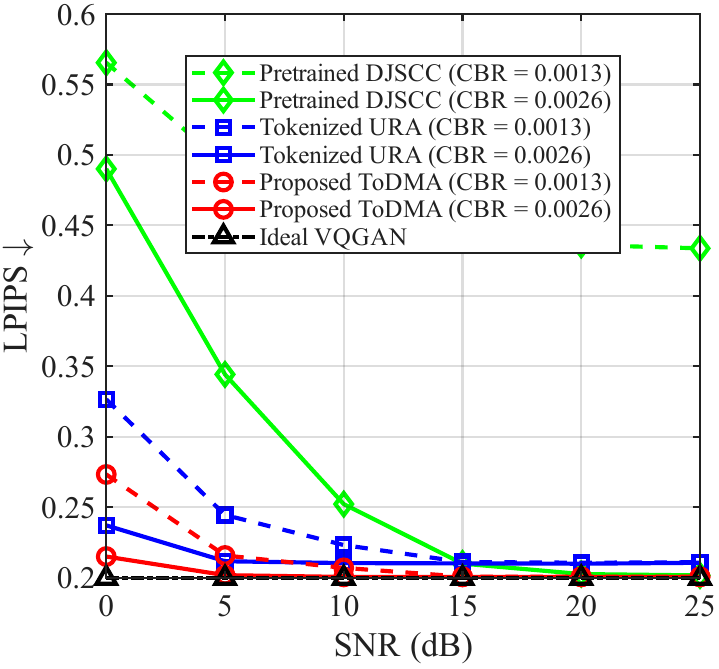}
        \label{fig:snr-c}
    }
    \subfloat[]{
        \includegraphics[width=0.24\textwidth]{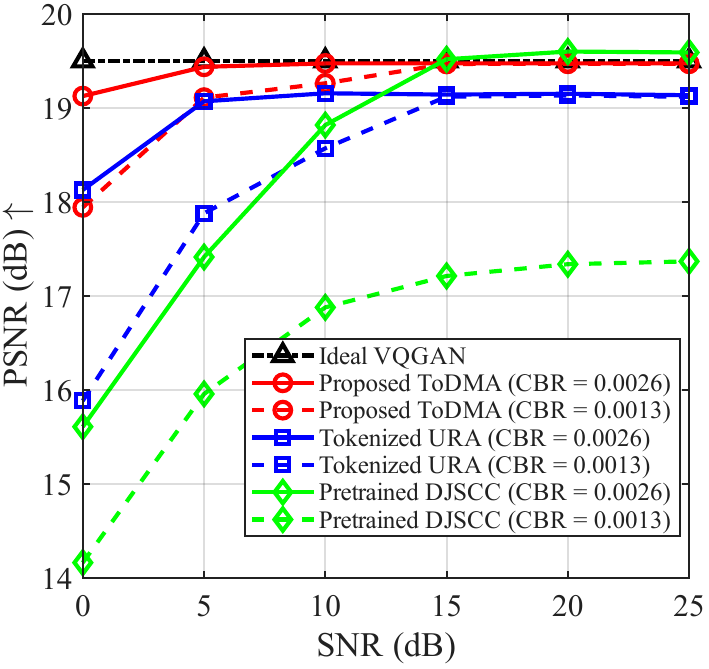}
        \label{fig:snr-d}
    } 
    \captionsetup{font={footnotesize, color={black}}, singlelinecheck=off, justification=raggedright, labelsep=period}
    \caption{Performance comparisons as a function of the channel SNR at $K=20$: (a) TER ($\downarrow$); (b) CLIP ($\uparrow$); (c) LPIPS ($\downarrow$); (d) PSNR ($\uparrow$). $\downarrow$: lower is better; $\uparrow$: higher is better.}
    \label{fig:snr_1x4}
\end{figure*}

\begin{figure*}[!t]
    \centering
    \captionsetup{font={footnotesize}, singlelinecheck=off, justification=centering, labelsep=period}
    \setlength{\tabcolsep}{1.5pt}

    \subfloat[]{
        \includegraphics[width=0.24\textwidth]{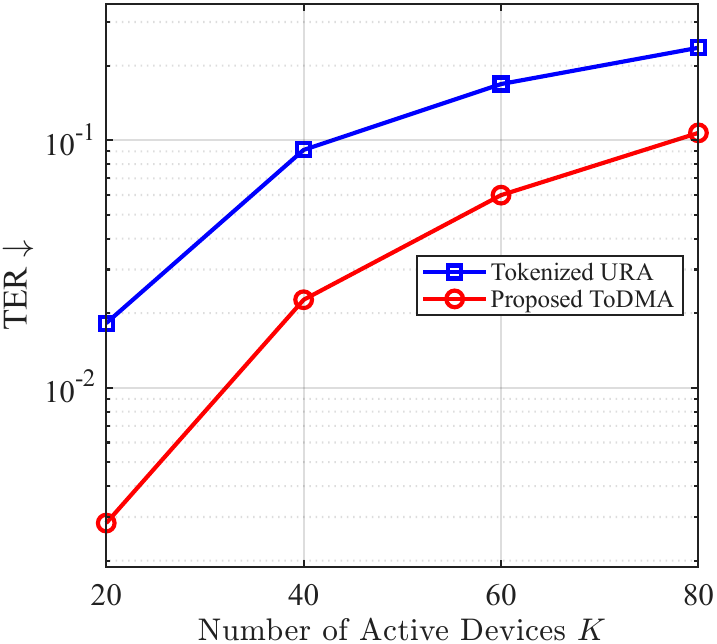}
        \label{fig:k-a}
    }
    \subfloat[]{
        \includegraphics[width=0.24\textwidth]{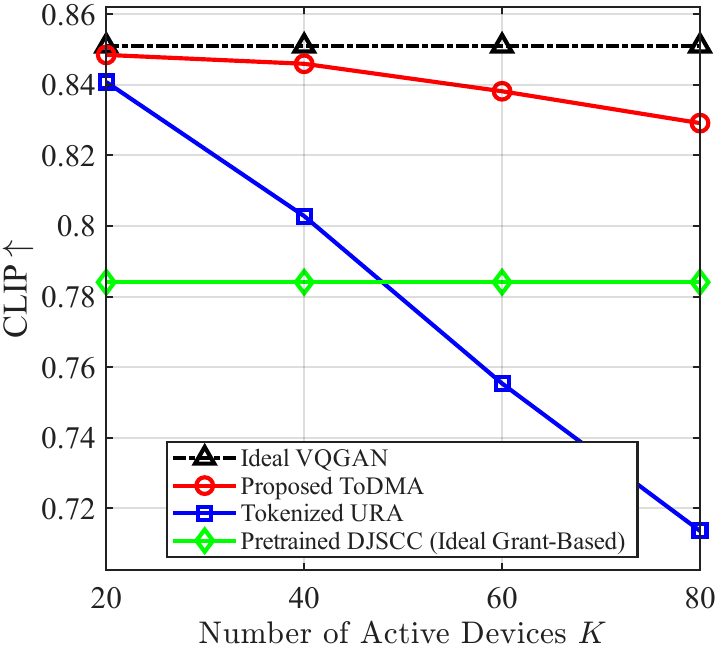}
        \label{fig:k-b}
    }
    \subfloat[]{
        \includegraphics[width=0.24\textwidth]{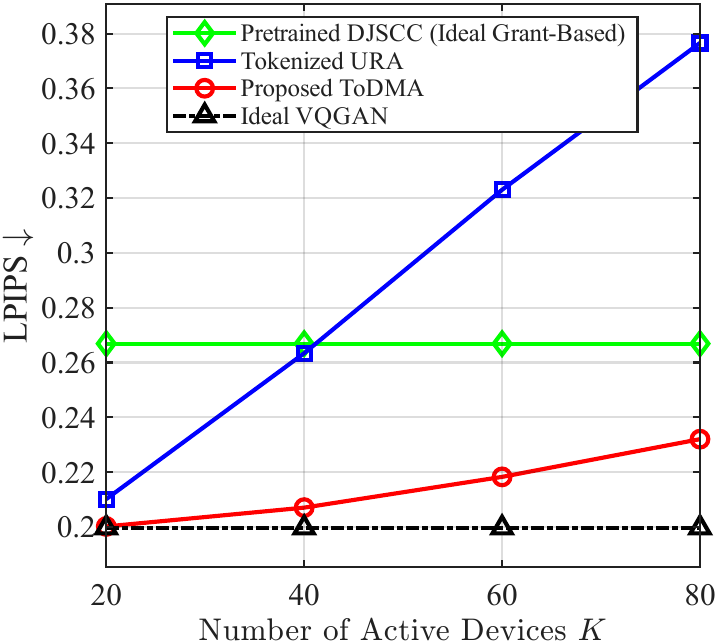}
        \label{fig:k-c}
    }
    \subfloat[]{
        \includegraphics[width=0.24\textwidth]{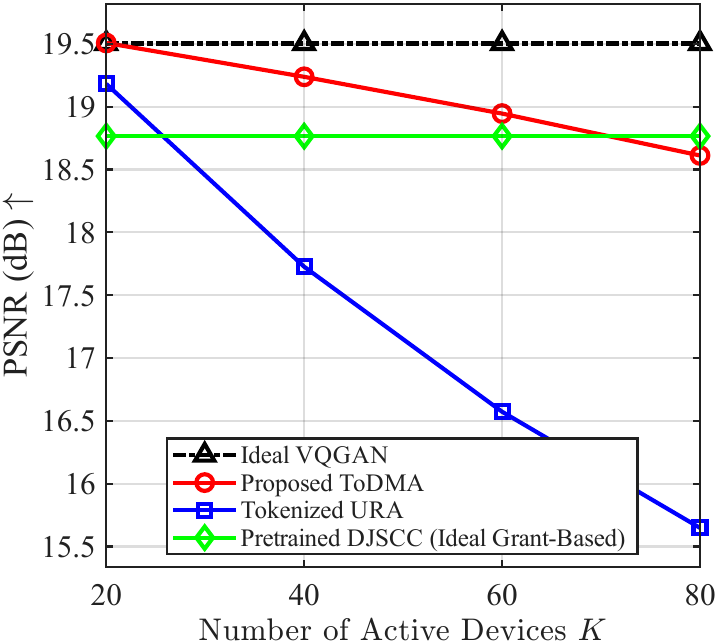}
        \label{fig:k-d}
    }
\captionsetup{font={footnotesize, color={black}}, singlelinecheck=off, justification=raggedright, labelsep=period}
    \caption{Performance comparisons as a function of the number of active devices $K$ at $\text{SNR}=10$\,dB: (a) TER ($\downarrow$); (b) CLIP ($\uparrow$); (c) LPIPS ($\downarrow$); (d) PSNR ($\uparrow$). {All communication schemes use a CBR of 0.0026.}}
    \label{fig:k_1x4}
\end{figure*}

{
Fig.~\ref{fig:snr_1x4} compares the TER and reconstruction quality of the different schemes versus SNR. For ToDMA, below $10$~dB, token errors arise from both collision ambiguity and physical-layer detection errors, whereas above $15$~dB, the residual TER is dominated by unresolved collisions. Nevertheless, ToDMA consistently achieves a lower TER than Tokenized URA over the entire considered SNR range, demonstrating the effectiveness of the proposed candidate assignment and context-aware masked prediction. The improved token reliability also translates into better reconstruction quality. With $\mathrm{CBR}=0.0013$, ToDMA substantially outperforms Pretrained DJSCC using the same CBR across all considered SNRs. It also outperforms Pretrained DJSCC with twice the channel bandwidth ratio, $\mathrm{CBR}=0.0026$, across all considered metrics at SNRs up to approximately $15$~dB. Beyond $15$~dB, the latter approaches the Ideal VQGAN benchmark in terms of CLIP and LPIPS and achieves a slightly higher PSNR, which can be attributed to its continuous latent representation. Overall, these results demonstrate the advantage of ToDMA under stringent communication-resource constraints.
}

 {
Fig.~\ref{fig:k_1x4} compares the performance of the different schemes as the number of active users increases from $K=20$ to $K=80$, with $\mathrm{CBR}=0.0026$ and $\mathrm{SNR}=10$~dB. As $K$ increases, more token collisions occur, causing the performance of Tokenized URA to degrade rapidly. In contrast, ToDMA uses candidate assignment and context-aware masked prediction to resolve a large fraction of the collided tokens. Consequently, its TER increases only moderately, while its reconstruction quality remains relatively stable. The performance advantage of ToDMA over Tokenized URA therefore becomes more pronounced as the network load increases, demonstrating the robustness of ToDMA to massive access. The Pretrained DJSCC baseline is evaluated under ideal GBRA, where access collisions and scheduling latency are neglected and orthogonal communication resources are assumed to be available for all active devices. Its reconstruction performance is therefore independent of $K$ in this comparison. However, supporting more active users through orthogonal resource allocation increases the communication latency, which will be discussed separately in the latency analysis.

}

\begin{figure}[t]
     \centering
     \includegraphics[width = 0.9 \columnwidth,keepaspectratio]{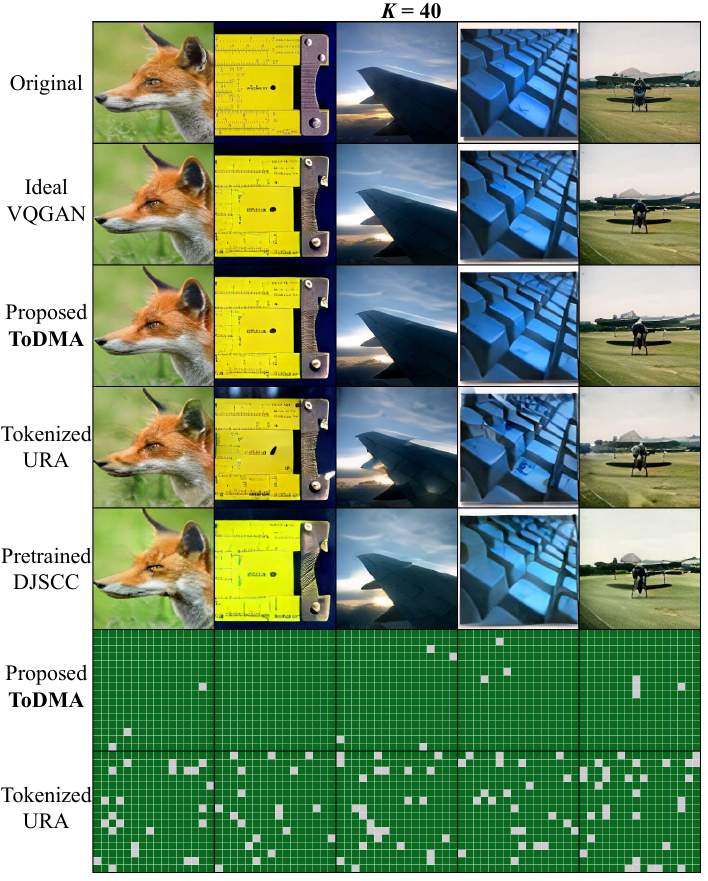}
     \captionsetup{font={footnotesize, color = {black}}, singlelinecheck = off, justification = justified,name={Fig.},labelsep=period}
     \caption{Visual illustrations for $K=40$. Rows from top to bottom: Original images and reconstructions using ``Ideal VQGAN'', ``ToDMA'', ``Tokenized URA'', and ``Pretrained JSCC''. The last two rows show token error maps, where green indicates correct tokens and gray indicates errors.}
     \label{Sim:Visual-1}
\end{figure}

\begin{figure}[t]
     \centering
     \includegraphics[width = 1 \columnwidth,keepaspectratio]{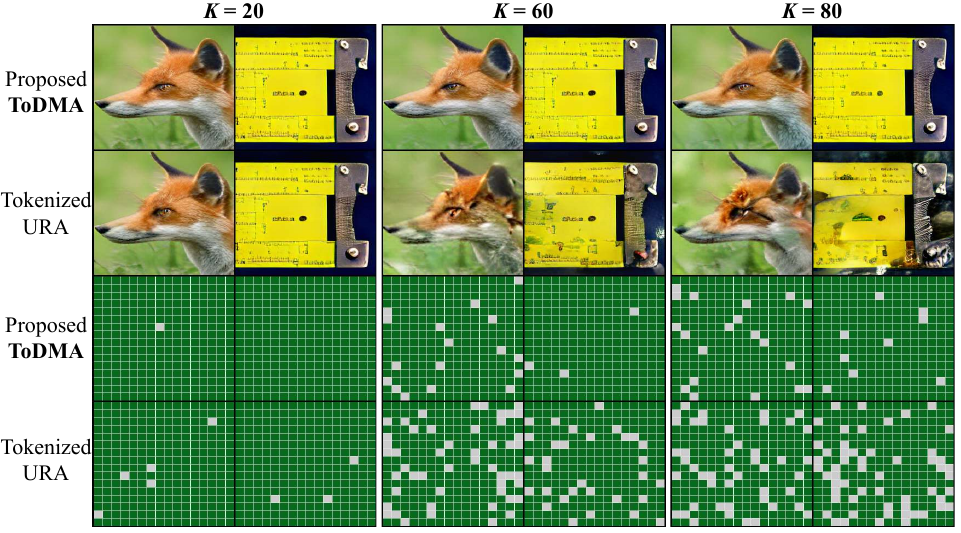}
     \captionsetup{font={footnotesize, color = {black}}, singlelinecheck = off, justification = justified,name={Fig.},labelsep=period}
     \caption{Visual illustrations for $K=\{20, 60, 80\}$. Rows from top to bottom: Reconstructions by ``ToDMA'' and ``Tokenized URA'', followed by their respective token error maps.} 
     \label{Sim:Visual-2}
\end{figure}

{
To provide a visual comparison, Figs.~\ref{Sim:Visual-1} and~\ref{Sim:Visual-2} present representative reconstruction results at $\mathrm{SNR}=10$~dB and $\mathrm{CBR}=0.0026$. For the token-based schemes, the corresponding $16\times16$ token-error maps are shown, where green and gray blocks denote correctly and incorrectly recovered tokens, respectively. As illustrated in Fig.~\ref{Sim:Visual-1} for $K=40$, Tokenized URA produces visible artifacts due to extensive token errors, whereas ToDMA reconstructs the image with quality close to the Ideal VQGAN benchmark. Its visual advantage over Pretrained DJSCC under ideal GBRA is also evident. Fig.~\ref{Sim:Visual-2} provides additional examples for $K=\{20,60,80\}$, showing that ToDMA preserves the main semantic and perceptual content across different network loads. The remaining errors may be further reduced through joint optimization of physical-layer detection and token prediction, stronger pretrained models, and iterative multi-round prediction. We further evaluate a pretrained image tokenizer\footnote{Model weight available at
\url{https://github.com/FoundationVision/LlamaGen}.} with a codebook size of
$Q=16384$. Under $K=20$,
$L=2K$, and $\mathrm{SNR}=25$~dB, ToDMA achieves a CLIP score of $0.9279$,
an LPIPS of $0.1335$, and a PSNR of $21.616$~dB, closely matching the ideal
performance of this tokenizer. This shows that
the performance of ToDMA can be further improved by adopting
a higher-quality tokenizer.
}

{
\subsection{ToDMA-Based Text Transmission}
\label{Sec:TextTransmission}

We further evaluate ToDMA for text transmission using the QUOTES500K dataset \cite{goel2018proposing}. In each Monte-Carlo trial, each active device constructs a text packet
comprising multiple randomly selected short sentences, with its token
length randomly selected from $\{120,\ldots,127\}$. All devices start
transmission after receiving a common beacon, while a device terminates
after sending its end-of-message token, without padding the remaining token
intervals. This variable-length setting requires no modification to the
receiver, since active-token detection and token assignment are performed
independently at each token position.


We use the WordPiece tokenizer and the pretrained ``BERT-base-uncased'' model \cite{devlin2018bert}. After removing unused special tokens and tokens absent from the QUOTES500K corpus, the communication codebook contains $Q=19534$ tokens and is shared by the transmitters and receiver. We measure the communication overhead by channel uses per token (CPT), where one channel use corresponds to one complex channel symbol. For ToDMA, we set $L=2K$, resulting in $\mathrm{CPT}\approx2$.

In addition to \emph{Tokenized URA}, we consider \emph{Text-DJSCC} and \emph{Token Arithmetic Coding} as baselines. \emph{Text-DJSCC} is trained on QUOTES500K following the semantic joint source and channel coding framework in \cite{xie2021deep}. It uses the embeddings generated by the same pretrained BERT model as ToDMA, followed by CNN-based dimensionality reduction and reconstruction, and is evaluated at $\mathrm{CPT}\approx2$ and $\mathrm{CPT}\approx4$. \emph{Arithmetic Coding} serves as a lossless digital baseline. Its static unigram probabilities are estimated directly from the text samples to be transmitted, resulting in an average compression rate of $9.48$ bits/token. The compressed bitstream is protected by a rate-$4/5$ low-density parity-check (LDPC) code and modulated using 64-ary quadrature amplitude modulation (64-QAM), yielding $\mathrm{CPT}\approx2$. No additional signaling overhead is included, providing a favorable digital baseline under the same communication-resource budget.

\begin{figure}[t]
\centering
\subfloat[]{
    \includegraphics[width=0.62\columnwidth]{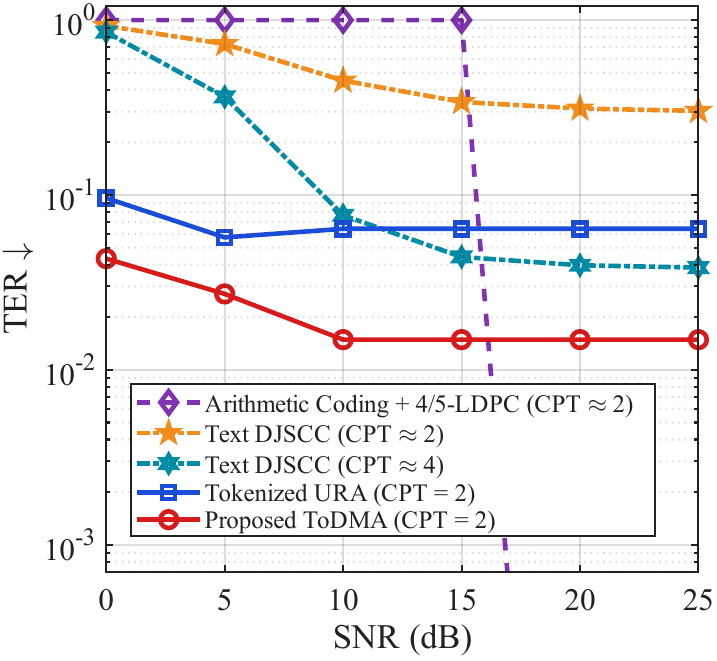}
    \label{FigTexta}
}
\hfill
\subfloat[]{
    \includegraphics[width=0.62\columnwidth]{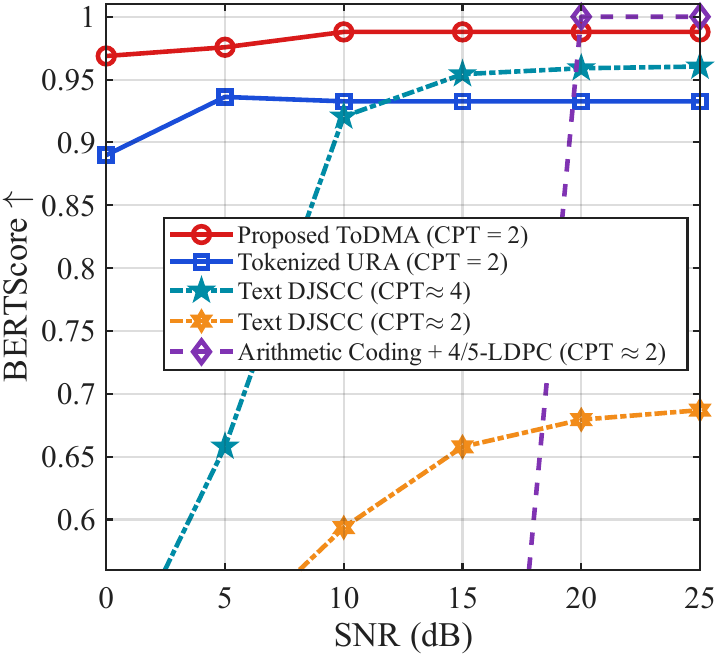}
    \label{FigTextb}
}
\captionsetup{font={footnotesize, color={black}}, singlelinecheck=off, justification=raggedright, labelsep=period}
\caption{Text transmission performance comparison at $K=20$.}
\label{Sim:Text}
\end{figure}

The performance is evaluated using TER and BERTScore \cite{zhang2020bertscore}, as shown in Fig.~\ref{Sim:Text}, where BERTScore measures semantic similarity by matching contextual token embeddings extracted using the pretrained BERT model. At $\mathrm{CPT}\approx2$, Text DJSCC performs substantially worse than ToDMA, whereas increasing its resource budget to $\mathrm{CPT}\approx4$ brings its performance close to that of ToDMA. This observation is consistent with Fig.~8 of \cite{xie2021deep}, which shows that increasing the number of transmitted symbols per word significantly improves text reconstruction. ToDMA also consistently outperforms Tokenized URA, demonstrating the effectiveness of candidate-restricted masked-token generation in resolving collided or missing tokens. Arithmetic Coding achieves error-free transmission above $20$~dB but exhibits a sharp cliff effect at $15$~dB, since residual bit errors can disrupt decoding and propagate through the sequence. For $Q=19534$, a fixed-length representation requires $15$ bits/token, while Arithmetic Coding reduces this to $9.48$ bits/token before channel coding and modulation. In contrast, ToDMA directly maps tokens to communication codewords, detects candidate tokens, and resolves ambiguity through contextual prediction. This avoids bit-level error propagation and enables graceful degradation under channel noise. Table~\ref{Tab:TextSample} presents a representative ToDMA text-transmission example for $K=15$ active devices and shows only the first ten token positions. Ambiguous positions are marked as [MASK], while tokens prefixed with ``\#\#'' denote WordPiece subword units.

}

{

\subsection{Cross-Modal Validation of Semantic Orthogonality}
\label{Sec:SO-sim}
To quantitatively validate the interpretation of semantic orthogonality in Section~\ref{Sec:SO}, we conduct collision-only Monte-Carlo experiments for both image and text transmission. The physical channel is excluded to isolate the semantic collision-resolution capability. For each value of $K$, the results are averaged over ten independent trials. Each image is represented by $256$ VQGAN tokens, while each text sample consists of $40$ WordPiece tokens. The candidate-set size and $\xi_{k,n}$ are evaluated and averaged over positions with $|\widetilde{\mathcal P}_n|>1$. Semantic quality is measured by CLIP similarity for the original and
reconstructed images and by BERTScore for the original and reconstructed texts.

As shown in Table~\ref{tab:so_validation}, increasing $K$ enlarges the candidate set and makes the collided tokens more difficult to distinguish based on their semantic context. Accordingly, as $K$ increases from $20$ to $80$, the average $\xi_{k,n}$ decreases from $0.8232$ to $0.7527$ for images and from $0.9044$ to $0.7677$ for text, accompanied by a consistent reduction in the token-recovery rate. These results verify that $\xi_{k,n}$ quantitatively captures the contextual separability of collision candidates and thus provides an operational measure of token-domain semantic orthogonality. The consistent behavior for image and text tokens further demonstrates the cross-modal applicability of this measure. In general, a larger
alphabet size $Q$ enlarges both the communication dictionary and the
unconstrained token-prediction space, making detection and contextual
recovery more challenging. Candidate restriction mitigates this issue:
although the image and text alphabets contain $Q=1024$ and $Q=19534$
tokens, respectively, their average candidate-set sizes in
Table~\ref{tab:so_validation} remain approximately 2--10.

Despite the reduced token-level separability under heavier collisions, the semantic-quality metrics degrade considerably more slowly. As $K$ increases from $20$ to $80$, the image token-recovery rate decreases from $0.8639$ to $0.5003$, whereas the CLIP similarity only decreases from $0.8258$ to $0.8215$. Similarly, the text token-recovery rate decreases from $0.9288$ to $0.7096$, while the BERTScore remains $0.9242$ at $K=80$. This behavior indicates that, even when the token selected by the pretrained model does not exactly match the original token, candidate-restricted prediction tends to select a contextually plausible alternative, thereby preserving the semantic consistency of the reconstructed content.

}

\begin{table*}[t]
\centering
\scriptsize
\caption{ToDMA-based text transmission with $K = 15$ active devices, displaying only the first ten tokens per device.}
\label{Tab:TextSample}
\begin{tabular}{|c|c|c|c|c|c|c|c|c|c|c|c|}
\hline
Positions & \tikz[baseline=(char.base)] \node[inner sep=0.3pt, minimum size=6pt, circle, draw, line width=0.3pt] (char) {0}; & \tikz[baseline=(char.base)] \node[inner sep=0.3pt, minimum size=6pt, circle, draw, line width=0.3pt] (char) {1}; & \tikz[baseline=(char.base)] \node[inner sep=0.3pt, minimum size=6pt, circle, draw, line width=0.3pt] (char) {2}; & \tikz[baseline=(char.base)] \node[inner sep=0.3pt, minimum size=6pt, circle, draw, line width=0.3pt] (char) {3}; & \tikz[baseline=(char.base)] \node[inner sep=0.3pt, minimum size=6pt, circle, draw, line width=0.3pt] (char) {4}; & \tikz[baseline=(char.base)] \node[inner sep=0.3pt, minimum size=6pt, circle, draw, line width=0.3pt] (char) {5}; & \tikz[baseline=(char.base)] \node[inner sep=0.3pt, minimum size=6pt, circle, draw, line width=0.3pt] (char) {6}; & \tikz[baseline=(char.base)] \node[inner sep=0.3pt, minimum size=6pt, circle, draw, line width=0.3pt] (char) {7}; & \tikz[baseline=(char.base)] \node[inner sep=0.3pt, minimum size=6pt, circle, draw, line width=0.3pt] (char) {8}; & \tikz[baseline=(char.base)] \node[inner sep=0.3pt, minimum size=6pt, circle, draw, line width=0.3pt] (char) {9}; & \multirow{2}*{\centering \makecell{Prediction\\ results}} \\ \cline{1-11}
\makecell{Candidate\\ token set} &
\makecell[l]{[`A',  `I']} & 
& & &
\makecell[l]{[`a',  `is']} & 
& & & & 
\makecell[l]{[`to',  `she']}  &   \\ \hline

Device 1 & Using & guilt & , & fear & or & any & other & negative & emotion & [MASK] & \makecell{\tikz[baseline=(char.base)] \node[inner sep=0.3pt, minimum size=6pt, circle, draw, line width=0.3pt] (char) {9};:`to'}\\ \hline
Device 2 & E & \#\#gna & \#\#ro & is & [MASK] & secret & known & to & everyone & but &  \makecell{\tikz[baseline=(char.base)] \node[inner sep=0.3pt, minimum size=6pt, circle, draw, line width=0.3pt] (char) {4};:`a'} \\ \hline
Device 3 & [MASK] & couldn & ' & t & think & of & anything & I & wanted & [MASK] &  \makecell{\tikz[baseline=(char.base)] \node[inner sep=0.3pt, minimum size=6pt, circle, draw, line width=0.3pt] (char) {0};:`I' \\\tikz[baseline=(char.base)] \node[inner sep=0.3pt, minimum size=6pt, circle, draw, line width=0.3pt] (char) {9};:`to'} \\ \hline
Device 4 & [MASK] & ’ & ve & always & believed & that & , & as & long & as &  \makecell{\tikz[baseline=(char.base)] \node[inner sep=0.3pt, minimum size=6pt, circle, draw, line width=0.3pt] (char) {0};:`I'} \\ \hline
Device 5 & Keep & me & up & till & five & because & all & your & stars & are &  \\ \hline
Device 6 & [MASK] & person & is & loved & for & the & risk & he & or & [MASK] &  \makecell{\tikz[baseline=(char.base)] \node[inner sep=0.3pt, minimum size=6pt, circle, draw, line width=0.3pt] (char) {0};:`A' \\\tikz[baseline=(char.base)] \node[inner sep=0.3pt, minimum size=6pt, circle, draw, line width=0.3pt] (char) {9};:`she'} \\ \hline
Device 7 & On & the & street & below & , & the & weather & is & calm & . &  \\ \hline
Device 8 & Men & , & discouraged & by & their & failure & to & accomplish & exactly & what &  \\ \hline
Device 9 & Any & preference & for & my & group & ' & s & interests & over & yours & \\ \hline
Device 10 & [MASK] & thing & about & poetry & [MASK] & , & It & takes & cuts & and &  \makecell{\tikz[baseline=(char.base)] \node[inner sep=0.3pt, minimum size=6pt, circle, draw, line width=0.3pt] (char) {0};:`A' \\\tikz[baseline=(char.base)] \node[inner sep=0.3pt, minimum size=6pt, circle, draw, line width=0.3pt] (char) {4};:`is'} \\ \hline
Device 11 & That & night & , & J & \#\#as & wore & the & only & dress & [MASK]  & \makecell{\tikz[baseline=(char.base)] \node[inner sep=0.3pt, minimum size=6pt, circle, draw, line width=0.3pt] (char) {9};:`she'} \\ \hline
Device 12 & In & Jesus & Christ & there & [MASK] & no & isolation & of & man & from & \makecell{\tikz[baseline=(char.base)] \node[inner sep=0.3pt, minimum size=6pt, circle, draw, line width=0.3pt] (char) {4};:`is'} \\ \hline
Device 13 & If & you & aren & ' & t & destroying & your & enemies & , & it  &  \\ \hline
Device 14 & He & pictured & himself & at & the & lake & , & on & a & house & \\ \hline
Device 15 & Su & \#\#pp & \#\#ose & that & [MASK] & great & commotion & arises & in & the  &  \makecell{\tikz[baseline=(char.base)] \node[inner sep=0.3pt, minimum size=6pt, circle, draw, line width=0.3pt] (char) {4};:`a'} \\ \hline
\end{tabular}
\end{table*}

\begin{table}[!t] \centering 
\caption{Validation of semantic orthogonality.} 
\label{tab:so_validation} 

\setlength{\tabcolsep}{3.2pt} \begin{tabular}{cccccc} \toprule Modality & $K$ & $\mathbb{E}[|\widetilde{\mathcal P}_n|]$ & $\mathbb{E}[\xi_{k,n}]$ & Token rec. & Sem. quality \\ \midrule Image & 20 & 2.14 & 0.8232 & 0.8639 & 0.8258 \\ 
Image & 40 & 3.23 & 0.8010 & 0.7522 & 0.8256 \\
Image & 60 & 5.76 & 0.7655 & 0.6105 & 0.8235 \\
Image & 80 & 9.66 & 0.7527 & 0.5003 & 0.8215 \\ \midrule 
Text & 20 & 2.41 & 0.9044 & 0.9288 & 0.9939 \\ 
Text & 40 & 3.96 & 0.8248 & 0.8411 & 0.9684 \\
Text & 60 & 6.61 & 0.7941 & 0.7642 & 0.9452 \\ 
Text & 80 & 9.33 & 0.7677 & 0.7096 & 0.9242 \\ \bottomrule 
\end{tabular}
\end{table} 

\begin{table}[t]
\centering
\caption{Computational complexity (GFLOPs/device) and measured runtime of the main AI modules.}
\label{tab:complexity}
\setlength{\tabcolsep}{1.8pt}
\renewcommand{\arraystretch}{1.15}

\begin{tabular}{lccccc}
\hline
Scheme &
Encoder &
Decoder &
\shortstack{Masked\\Prediction} &
Total &
\shortstack{Wall-clock\\(ms/device)}
\\ \hline

\multicolumn{6}{c}{\textit{Image transmission (256 tokens/image)}} \\ \hline
Pretrained DJSCC
& 69.32
& 126.43
& N/A
& 195.87
& 8.80
\\

Tokenized URA
& 69.32
& 126.43
& N/A
& 195.75
& 8.87
\\

ToDMA
& 69.32
& 126.43
& 46.95
& 242.70
& 12.01
\\ \hline

\multicolumn{6}{c}{\textit{Text transmission (128 tokens/sequence)}} \\ \hline
Text DJSCC
& 11.17
& 3.08
& N/A
& 14.25
& 0.509
\\

Tokenized URA
& N/A
& N/A
& N/A
& N/A
& N/A
\\

ToDMA
& N/A
& N/A
& 14.25
& 14.25
& 0.488
\\ \hline
\end{tabular}

\begin{minipage}{0.99\columnwidth}
\scriptsize
All results are measured on a single NVIDIA A100 80\,GB PCIe GPU.
Following the {fvcore} convention, one multiply-accumulate
operation is counted as one FLOP.
\end{minipage}

\end{table}
\subsection{Latency Analysis}
\label{subsec:latency}
{We separately evaluate the communication latency and computational latency of the considered schemes. To provide a unified comparison for both image and text transmission, we use the same payload channel-use budget.} We consider a standard 5G New Radio (NR) system operating with a 10\,MHz bandwidth, which corresponds to a typical subcarrier spacing of 15 kHz. The time–frequency resources are organized into orthogonal frequency-division multiplexing (OFDM) symbols and slots. Specifically, each OFDM symbol carries 624 complex symbols, and the symbol duration including cyclic prefix is approximately 71.4 \textmu s. A slot consists of 14 OFDM symbols, resulting in a slot duration of 1 ms. 

\begin{figure}[t]
     \centering
     \includegraphics[width = 0.63\columnwidth,keepaspectratio]{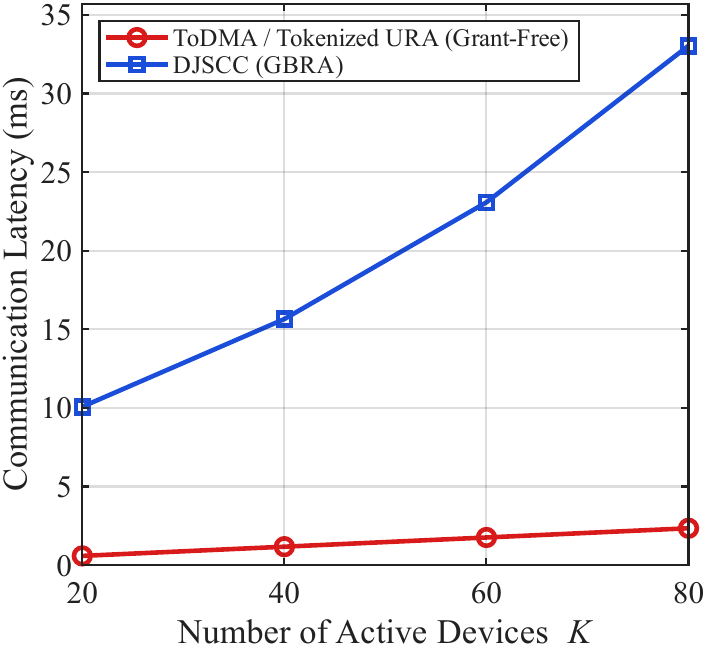}
     \captionsetup{font={footnotesize, color = {black}}, singlelinecheck = off, justification = justified,name={Fig.},labelsep=period}
     \caption{Communication latency versus the number of active devices with 256 transmitted symbols per device.}
     \label{Sim:Latency}
\end{figure}

For a fair comparison, each active device is allocated 256 complex symbols. Hence, both ToDMA with $L=2K$ and the orthogonal grant-based DJSCC baseline require a payload transmission latency of $\frac{256K}{624}\times71.4~\text{\textmu s}$. For the ToDMA and the Tokenized URA baseline, we reasonably assume that preconfigured grant-free resources are available, and thus no additional grant-acquisition delay is considered. In contrast, DJSCC relies on GBRA before payload transmission. Following \cite{LatencyEst}, the contention-free grant-acquisition procedure incurs a minimum latency of $T_{\min}=6$~ms, including the scheduling request (1~ms), next-generation NodeB (gNB) processing (2~ms), downlink-grant transmission (1~ms), and device preparation (2~ms).
{
However, this 6~ms latency represents a single-attempt lower bound and does not account for random-access contention. In each random-access opportunity, every device independently selects one of the $N_{\mathrm{RA}}$ available contention-based preambles. If multiple devices select the same random-access preamble, they may receive the same random-access response and use the same uplink grant for the scheduled uplink transmission, resulting in contention-resolution failure and another access attempt~\cite{LatencyEst,3GPP38321}. Under independent and uniform preamble selection, the success probability of a tagged device is $(1-1/N_{\mathrm{RA}})^{K-1}$. Using a geometric-attempt approximation, the access latency of the GBRA baseline is given by \begin{equation} T_{\mathrm{access}}^{\mathrm{GBRA}}(K) \approx T_{\min} + \left[ \left(1-\frac{1}{N_{\mathrm{RA}}}\right)^{-(K-1)} -1 \right]T_{\mathrm{ret}}, \label{eq:contention_aware_communication_latency} \end{equation} where $T_{\mathrm{ret}}$ is the average delay before a repeated access attempt. We set $T_{\mathrm{ret}}=10$~ms, corresponding to the average random backoff under a 20~ms backoff parameter specified in \cite{3GPP38321}. Other delays, such as contention-resolution waiting, waiting for the next random-access opportunity, control-channel blocking, and scheduling queues, are omitted, making this estimate favorable to the GBRA baseline. Common synchronization and system-configuration overheads are omitted for all schemes. Based on these settings, Fig.~\ref{Sim:Latency} compares the communication latency of the considered schemes as the number of active devices increases.
}

{
The computational complexity of the main AI modules is summarized in
Table~\ref{tab:complexity}. The AMP detector is omitted here, as its
complexity has been analyzed in the preceding section. We also report the
measured wall-clock time on a single NVIDIA A100 80,GB PCIe GPU with
$K=20$, covering the dominant transmitter- and receiver-side AI modules.

For image transmission, the transmitter-side VQGAN encoder requires
approximately 3.35~ms/device and is included in the reported runtime.
Pretrained DJSCC and Tokenized URA have similar complexity and runtimes of
approximately 8.8~ms/device, since both mainly use the same VQGAN encoder
and decoder. ToDMA additionally performs receiver-side masked-token
prediction, increasing its runtime to 12.01~ms/device while enabling
context-aware token recovery and improved reconstruction quality.

For text transmission, WordPiece tokenization and codeword indexing in
ToDMA require no neural-network forward pass, whereas Text DJSCC employs a
BERT encoder at the transmitter. The two schemes have essentially the same
total AI complexity and comparable runtimes of 0.488 and 0.509~ms/device,
respectively, but place their main computation at the receiver and
transmitter, respectively. Thus, ToDMA's performance gain arises from
non-orthogonal token transmission and candidate-restricted prediction,
rather than a larger computation budget. Moreover, Tokenized URA
outperforms Text DJSCC under some settings in Fig.~\ref{Sim:Text}, further
demonstrating the benefit of token-domain multiple access.

Since computational runtime depends on the hardware platform, batch size, and implementation, it is reported separately from communication latency. Overall, ToDMA introduces additional receiver-side computation to improve semantic recovery. Meanwhile, its performance gains over DJSCC cannot be attributed merely to a larger neural-network complexity.

}

\balance
\section{Conclusions}

This work investigated massive uplink transmission of tokenized source representations, which are emerging as a common interface in large-model-enabled intelligent systems. In the considered semantic multiple access setting, multimodal source signals are represented as compact discrete tokens drawn from pretrained codebooks. Unlike conventional bit streams, token sequences possess a discrete codebook structure and contextual dependencies that can be exploited in multiple access design. ToDMA leverages these properties by associating token indices with shared random access codewords, enabling massive uncoordinated devices to transmit token sequences in a grant-free and non-orthogonal manner. At the receiver, compressed sensing-based token detection, CSI consistency-based token assignment, and candidate-restricted contextual prediction are combined to recover the superposed token sequences. We further introduced an entropy-based interpretation of token-domain semantic orthogonality as an operational measure of the contextual separability among physically ambiguous token candidates.

Extensive experiments on image and text transmission demonstrate that ToDMA better preserves token-level and semantic consistency under a low channel use budget. The results show that its performance gain does not simply come from a stronger neural backbone or a larger computation budget. Instead, it arises from the joint use of non-orthogonal token-domain transmission, physical-layer candidate restriction, and contextual token recovery. Comparisons with pretrained DJSCC, context-unaware tokenized URA, and conventional digital transmission confirm the bandwidth efficiency and robustness of the proposed framework. The latency and complexity analyzes further show that ToDMA trades moderate receiver-side AI computation for reduced signaling overhead and improved token recovery efficiency.

More broadly, ToDMA suggests that tokens should not be treated merely as source symbols to be converted into bits. Since tokens are drawn from pretrained codebooks and exhibit contextual predictability, their structure can be exploited directly in the design of the communication system. Future research may proceed along three directions. At the algorithmic level, joint token detection and context prediction can be developed to enable iterative refinement between physical-layer recovery and semantic inference. At the system level, channel adaptation, semantic-aware modulation-codebook design, and consistent token interpretation across heterogeneous devices are important for practical design. At the application level, task-oriented token communication can further incorporate token importance, multimodal token interactions, and downstream AI objectives into token selection and resource allocation.

\bibliographystyle{ieeetr}
\bibliography{references}

\end{document}